\documentclass[12pt]{article}

\usepackage{graphicx}
\usepackage{epsfig,cite}
\usepackage{amsmath}
\usepackage{amssymb}
\usepackage{latexsym,url}

\def\jhep #1#2#3 {{JHEP} {\bf#1} (#2) #3}
\def\plb #1 #2 #3 {{Phys.~Lett.} {\bf B#1} (#2) #3}
\def\npb #1 #2 #3 {{Nucl.~Phys.} {\bf B#1} (#2) #3}
\def\epjc #1#2#3 {{Eur.~Phys.~J.} {\bf C#1} (#2) #3}
\def\epjd #1#2#3 {{Eur.~Phys.~J.} {\bf D#1} (#2) #3}
\def\zpc #1 #2 #3 {{Z.~Phys.} {\bf C#1} (#2) #3}
\def\jpg #1 #2 #3 {{J.~Phys.} {\bf G#1} (#2) #3}
\def\prd #1#2#3 {{Phys.~Rev.} {\bf D#1} (#2) #3}
\def\prep #1 #2 #3 {{Phys.~Rep.} {\bf#1} (#2) #3}
\def\prl #1 #2 #3 {{Phys.~Rev.~Lett.} {\bf#1} (#2) #3}
\def\mpl #1 #2 #3 {{Mod.~Phys.~Lett.} {\bf#1} (#2) #3}
\def\rmp #1 #2 #3 {{Rev. Mod. Phys.} {\bf#1} (#2) #3}
\def\cpc #1 #2 #3 {{Comp. Phys. Commun.} {\bf#1} (#2) #3}
\def\sjnp #1 #2 #3 {{Sov. J. Nucl. Phys.} {\bf#1} (#2) #3}
\def\xx #1 #2 #3 {{\bf#1}, (#2) #3}
\def\hepph #1 {{\tt hep-ph/#1}}

\newcommand{\be}{\begin{equation}}
\newcommand{\lp}{\left(}
\newcommand{\rp}{\right)}
\newcommand{\la}{\left\langle}
\newcommand{\ra}{\right\rangle}

\newcommand{\ee}{\end{equation}}
\newcommand{\bea}{\begin{eqnarray}}
\newcommand{\eea}{\end{eqnarray}}

\newcommand{\smallz}{{\scriptscriptstyle Z}} %
\newcommand{\mz}{m_\smallz}
\newcommand{\smallw}{{\scriptscriptstyle W}}
\newcommand{\mw}{m_\smallw} 
\newcommand{\gw}{\Gamma_{\smallw}}

\newcommand{\sdw}{\sin^2\theta_{\smallw}}

\newcommand{\smallh}{{\scriptscriptstyle H}}
\newcommand{\mh}{m_\smallh}

\newcommand{\oa}{${\cal O}(\alpha)~$}

\begin{document}

\newenvironment{appendletterA}
 {
  \typeout{ Starting Appendix \thesection }
  \setcounter{section}{0}
  \setcounter{equation}{0}
  \renewcommand{\theequation}{A\arabic{equation}}
 }{
  \typeout{Appendix done}
 }

\begin{titlepage}
\nopagebreak

{\flushright{
        \begin{minipage}{5cm}
         IFUM-961/FT \\ 
        \end{minipage}        }}

\renewcommand{\thefootnote}{\fnsymbol{footnote}}
\vskip 1.5cm
\begin{center}
\boldmath
{\Large\bf The impact of PDF uncertainties }\\[9pt]
{\Large\bf on the measurement of the W boson mass }\\[9pt]
{\Large\bf at the Tevatron and the LHC}\unboldmath
\vskip 1.cm
{\large G.~Bozzi\footnote{Email: Giuseppe.Bozzi@mi.infn.it},
        J.~Rojo\footnote{Email: Juan.Rojo@mi.infn.it} and
        A.~Vicini\footnote{Email: Alessandro.Vicini@mi.infn.it}
}
\vskip .2cm
{\it Universit\`a degli Studi di Milano and
INFN, Sezione di Milano,\\
Via Celoria 16, 
I-20133 Milano, Italy} \\[5mm]

\end{center}

\begin{abstract}
We study at a quantitative level the impact of the uncertainties on the
value of the $W$ boson mass measured at hadron colliders due to: {\it
  i}) the proton parton distribution functions (PDFs), {\it ii}) the
value of the strong coupling constant $\alpha_s$ and {\it iii}) the
value of the charm mass used in the PDF determination. The value of the
$W$ boson mass is extracted, by means of  a template fit technique, from
the lepton-pair transverse mass distribution measured in the charged
current Drell-Yan process. We study the determination of $\mw$ at the
Tevatron and at the LHC with 7 and 14 TeV of center-of-mass energy in a
realistic experimental setup. The analysis has been done at the
Born level using the
event generator HORACE and at NLO-QCD using the event generators
DYNNLO and ResBos. We consider the three global
PDF sets, CTEQ6.6,
MSTW2008 and NNPDF2.1. We estimate that the total 
PDF uncertainty on $\mw$ is below 10 MeV both at the Tevatron and at the
LHC for all energies and final states. We conclude
that PDF uncertainties do not challenge a measurement of the 
$W$ boson mass at the level of 10 MeV accuracy.
\end{abstract}

\vfill
\end{titlepage}    

\setcounter{footnote}{0}
\tableofcontents

\section{Introduction}
\label{sec:intro}
The measurement of the $W$ boson mass represents a very important test of
the Standard Model and of its extensions, like e.g. the MSSM, and
provides indirect bounds on the mass of the Higgs 
boson~\cite{ewwg,Hoecker:2009gd,Heinemeyer:2007bw}. 
This measurement has reached a very high level of
accuracy: the current world average is $\mw=80.398\pm0.023$ 
GeV~\cite{:2009nu} and the best single experiment measurements have been
obtained by D0~\cite{Abazov:2009cp} and 
CDF~\cite{Aaltonen:2007ps,:2007ypa} at the Fermilab Tevatron with
$\mw=80.401\pm 0.043$~GeV and $\mw=80.413\pm 0.048$~GeV respectively. The
prospects for the combined measurements at the end of the Tevatron run,
with 4 fb$^{-1}$  of total collected luminosity, are of a final error of
roughly 15 MeV~\cite{Zhu:2009fx}. The prospects for the measurement at
the CERN LHC are at the level of 15 MeV, or even 10
MeV~\cite{cmsnote,atlasnote}.
At this level of accuracy it becomes necessary to
quantify in detail the various sources of theoretical 
uncertainties that contribute to the final
systematic error. 

The mass of the $W$ boson is measured at hadron colliders in the charged
current (CC) Drell-Yan (DY) process by studying the charged lepton
transverse momentum $p_t^l$ distribution, the missing transverse momentum
$p_t^{\nu}$
distribution or the lepton pair transverse mass distribution, defined
as
\begin{equation}
M_{\perp}^{W}=\sqrt{2p_t^lp_t^{\nu}\left( 1-\cos\lp
    \phi^l-\phi^{\nu}\rp\right)}\ ,
\end{equation}
where the neutrino four-momentum $p_t^{\nu}$
and angle $\phi^{\nu}$ are inferred from the transverse momentum
imbalance in the event.
The mass
of the $W$ boson is obtained by fitting the experimental distributions
with the corresponding theoretical predictions, where  $\mw$ is
kept as a free parameter. 

\begin{figure}[t]
\begin{center}
\includegraphics[height=60mm]{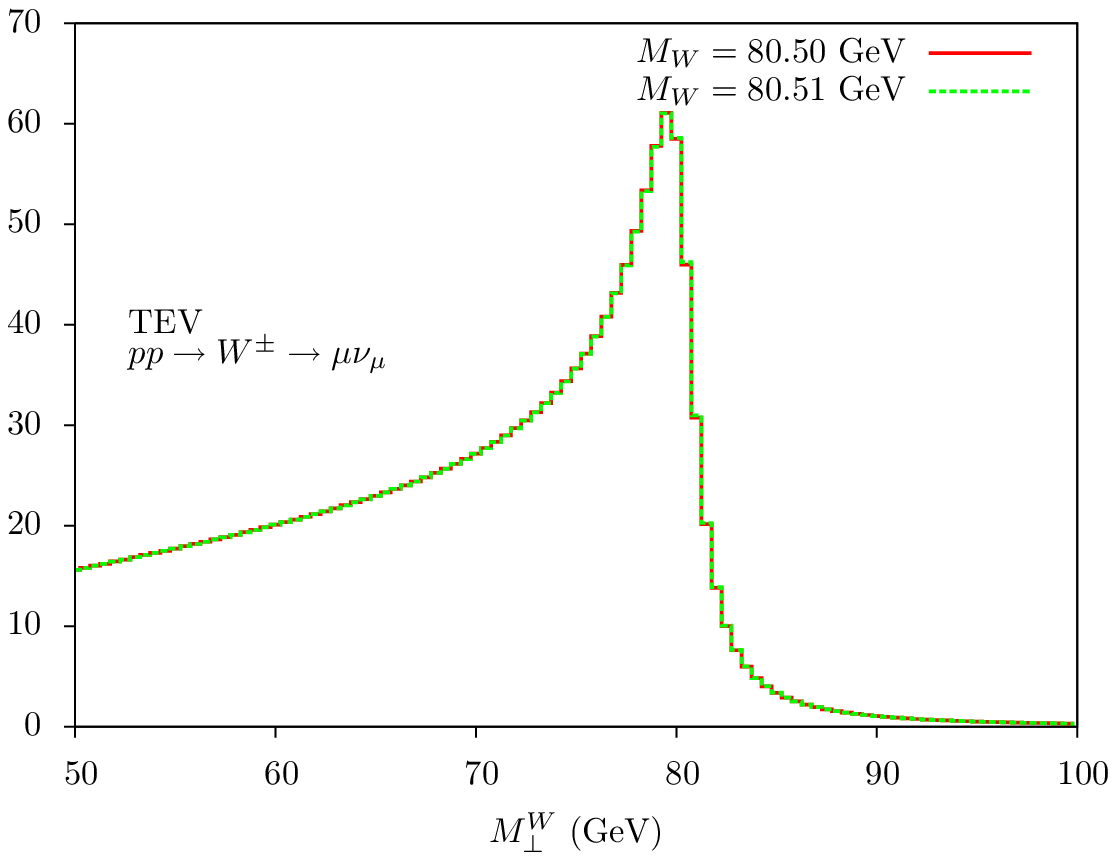}~
\includegraphics[height=60mm]{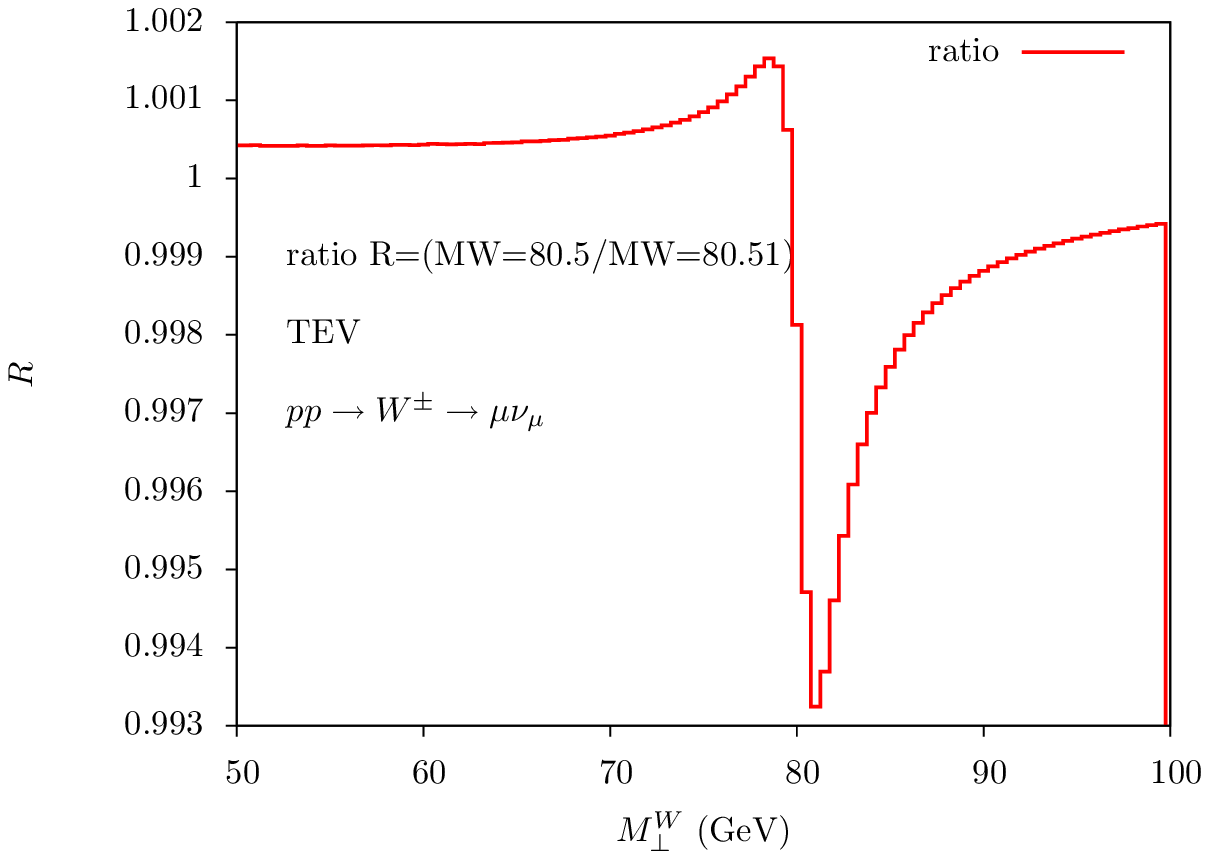}
\end{center}
\caption{\small Left plot: the transverse mass distributions at the
  Tevatron at Born level obtained with two values of $\mw$ which differ
  by 10 MeV. Right plot: the bin by bin ratio of these two distributions.} 
\label{ratiotemplates}
\end{figure}

A measurement of $\mw$ at the 10 MeV level is not only a very ambitious
goal from the experimental side, but it is also very
challenging from the theoretical point of view due to the careful
modelling of the production mechanism that is required.
We can illustrate these difficulties with the following
example. It is  known that the result of a fit of $\mw$ to
a given theory template is very
sensitive to the shape of the distributions.  
In Fig.~\ref{ratiotemplates} we consider two transverse mass distributions
at the Born level obtained
with two values of $\mw$ which differ by 10 MeV.
If one takes the ratio bin by bin of the histograms, one sees that
a small shift of 10 MeV in $\mw$ induces a
non trivial distortion of the shape at the permille level. Therefore, if
we aim at measuring $\mw$ at the 10-20 MeV level, we should, from the
theoretical side, have the control on all the perturbative and
non-perturbative corrections which can change the shape of the relevant
kinematic distributions at this level of precision.

On the other hand,
the total integrated cross-section is not significantly affected 
by changing $\mw$. As shown in Table~\ref{table:sigmatot}, a shift by 10
MeV of $\mw$ yields a change of the cross section at the 0.04\% level.
Thus, it is important to disentangle the normalization effects, which
are very weakly related to the precise value of $\mw$, from the effects
that modify instead the shape of the distributions, which have a larger
impact on the measurement of $\mw$. 

\begin{table}[t]
\begin{center}
\small
\begin{tabular}{|c|c|c|c|c|c|c|}
\hline
$\mw$ (GeV) & 80.368 & 80.378 & 80.388 & 80.398 & 80.408 & 80.418  \\
\hline
 $\sigma_{tot}(\mw)$ (pb) & 368.72 & 368.87 & 369.03 &
  369.17 & 369.32 & 369.46 \\
\hline
$\lp \sigma_{tot}^{i+1}-\sigma_{tot}^{i}\rp/\sigma_{tot}^{i}$ & & 0.04\%
& 0.04\% & 0.04\% & 0.04\% & 0.04\%  \\
\hline
\end{tabular}
\caption{\small Cross sections within acceptance cuts, at Born level, as a
  function of $\mw$. We also show the percentage difference between pairs
  of cross sections that differ by 10 MeV.} 
\label{table:sigmatot}
\end{center}
\end{table}

The Drell-Yan cross-section is given by the convolution of the parton
distribution functions (PDFs) of
the two incoming hadrons with the partonic cross-section. The crucial
role of QCD corrections to the partonic processes has been widely
discussed in the literature~\cite{Anastasiou:2003ds,Catani:2009sm}. 
The very important role of the \oa EW
corrections in the precision study of the CC DY process is also well
known (for a complete list of references see~\cite{Balossini:2009sa}).
It is the aim of the present paper is to study three different
sources of uncertainty related to the PDFs and their impact on the
measurement of $\mw$:
\begin{enumerate}
\item
the PDFs are affected by uncertainties due to the error of the
experimental data from which they are extracted, as well as by
theoretical uncertainties like the non-perturbative functional
form parametrization. These uncertainties
affect the prediction of the DY observables and, in turn, the extraction
of the value of $\mw$. Moreover in some cases the central values
obtained from different PDF sets differ more than the nominal
PDF uncertainties: we need to account for this by considering more than
one PDF set;
\item
the NLO-QCD  corrections sizably modify the Born level lepton transverse
momentum distribution and, more moderately, also the transverse mass
distribution. The precise effect of these corrections depends on the
value of the strong coupling constant, which is ultimately correlated
with the PDFs and with their evolution. Therefore also the precise value
of $\alpha_s$ should be taken into account
in a precision determination of $\mw$;
\item the PDFs depend on the value of the heavy quark masses
$m_c$ and $m_b$ due to two different reasons: the first one is
the fact that $\mathcal{O}\lp m_c^2/Q^2\rp$ terms have a 
non-negligible impact on PDF fits, and the second that 
heavy quark PDFs are obtained by assuming them to vanish at threshold,
and then to be generated by perturbative evolution. For these reasons, the
value of the $m_c$ used in the PDF determination has an impact on  the 
kinematic distributions from which $\mw$ is extracted, and tuhs must be
accounted for. The value of $m_b$ on the other hand does not affect $W$
production due to the smallness of b-initiated contributions.
\end{enumerate}

Uncertainties related to PDFs are known to be an important
component of the total systematic error in the determination of
$\mw$ at hadron colliders.
In the most recent CDF and D0 measurements PDF uncertainties
are estimated to be between 10 and 13 MeV~\cite{:2009nu}. 
Ref.~\cite{atlasnote} estimates PDF errors in $\mw$ prior to LHC data to be
$\sim$25 MeV, decreasing at the few MeV level once the constraints
from LHC processes are taken into account. 
On the other hand, there are claims~\cite{Krasny:2010vd}
that with the
current knowledge of PDFs a determination of $\mw$ with
a precision $\Delta \mw\le 10$ MeV is far from being possible.
In this paper we want to revisit the impact of PDFs and related
uncertainties on the determination of $\mw$ at the Tevatron
and the LHC, considering the most updated global PDF sets
and related theoretical uncertainties,
like the values of $\alpha_s$ and $m_c$.

The paper is organized as follows. In Sect.~\ref{fitting} we present
the general strategy that we will follow to estimate the shifts of the
measured value of $\mw$  induced by PDF uncertainties. 
In Sect.~\ref{resultsmtw} we present the numerical results of our
analysis for the transverse mass distribution and
in Sect.~\ref{resultsmw} the results for the PDF impact on the
determination of $\mw$. In Sect.~\ref{pdflhc} we explore
the improvements on PDF uncertainties for the
determination of the $W$ mass provided by LHC data
and in  Sect.~\ref{concl} we draw our conclusions. 


\section{The determination of $\mw$: general strategy}
\label{fitting}

In this section we present the general strategy that we adopt
to estimate the impact of PDF uncertainties in the determination of
$\mw$ at hadron colliders. First of all, we introduce the fitting
procedure and its validation. Then we discuss the event generators
and settings adopted to compute the theoretical distributions.
Finally, we discuss the PDF sets that are considered in this study
together with related sources of theoretical uncertainty.

\subsection{The fitting procedure and its validation}

We consider in the present study differential distributions in Charged
Current Drell-Yan production generated
with different PDF sets and we treat them as samples of pseudodata. The
Montecarlo error on each bin is taken in the statistical analysis as the
error affecting the pseudodata. The pseudodata are generated with a
given common nominal value of $\mw$ called $\mw^0$, which is taken 
to be $\mw^0=80.398$ GeV, the current world average.

\begin{figure}[t]
\begin{center}
\includegraphics[width=0.90\textwidth]{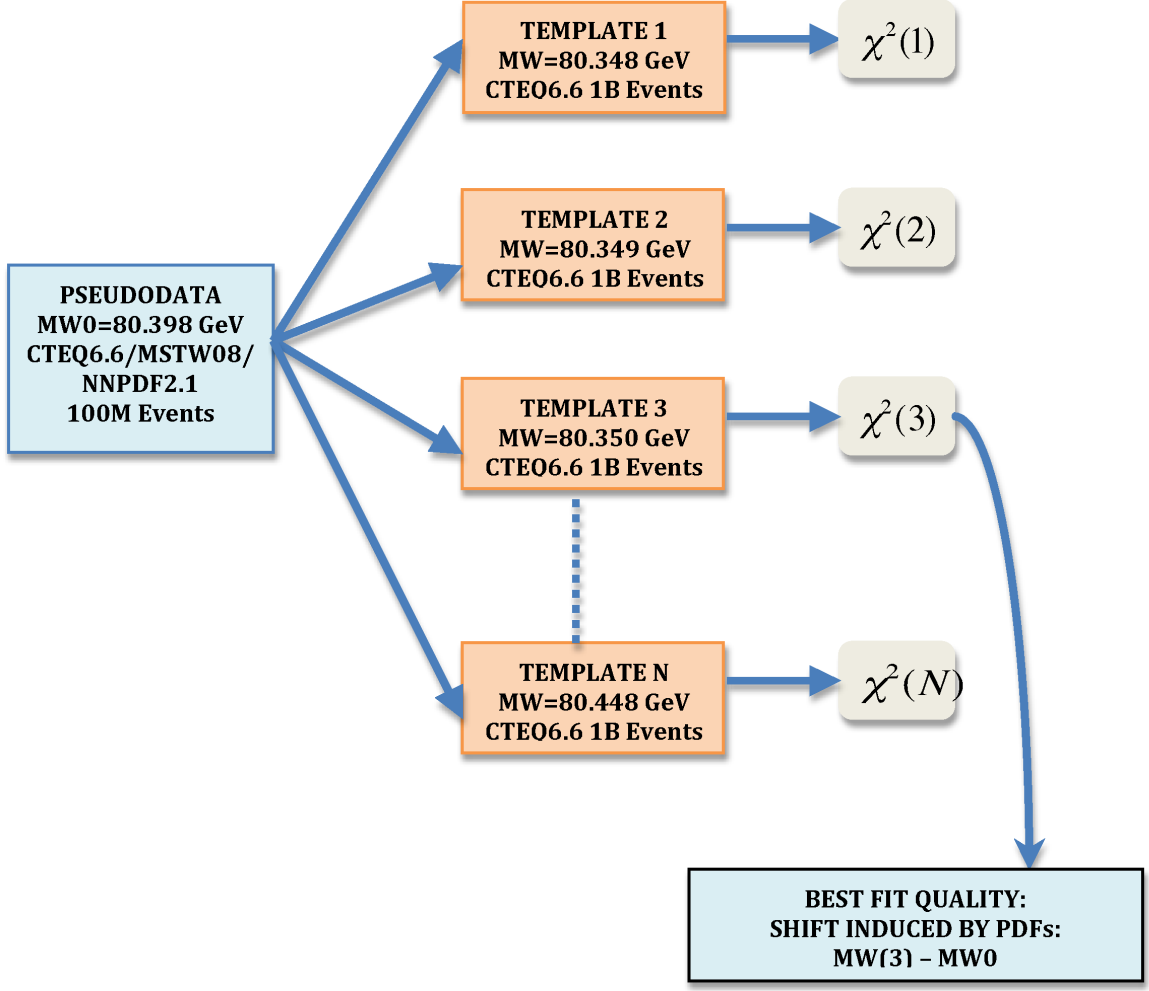}~
\end{center}
\caption{\small Flowchart that summarizes the procedure used to
determine the shift in $\mw$ induced by any given PDF set. More details
are provided in the text.} 
\label{fig:flowchart}
\end{figure}

The general fitting strategy is summarized in Fig.~\ref{fig:flowchart}.
First of all, we generate the templates for a given fixed PDF set,
in this case the central set of CTEQ6.6, and for different values
of $\mw$, with very high statistics, 1B events at
Born level. Then for each member of the
PDF sets considered, including the error PDF sets, we generate
pseudo-data with fixed $\mw^0=80.398$ GeV using exactly the same 
event generator as for the templates, with lower
statistics, 100M events at Born level. Then we compute the $\chi^2$
between the pseudo-data and each of the templates: the template with
best $\chi^2$ provides the information on which is the shift in 
$\mw$ induced by this particular PDF set. As expected for
consistency when pseudo data is generated with the central CTEQ6.6
set, we get  $\chi^2\sim 1$ for the selected template with $\mw=\mw^0$

The templates have been computed for 100 (at Born level)
and 20 (at the NLO-QCD level) different values of the $W$ mass.
The range for these templates has been taken to be $80.398\pm 0.050$ MeV
at Born level and $80.398\pm 0.036$ at NLO-QCD. 
We compare each template with the
pseudodata and compute the reduced $\chi^2$ function, defined as
\be
\chi_j^2=
\frac{1}{N_{\rm bins}}
\sum_{i=1}^{N_{\rm bins}}
\frac{\left(\mathcal{O}_i^{j}-\mathcal{O}_i^{\rm
      data}\right)^2}{(\sigma_i^{\rm data})^2 }~~~~~~j=1,\dots,N_{\rm
  templates}
\label{chi2j}
\ee
where $\mathcal{O}_i$ is the value of the 
$i-$th bin of the distribution $\mathcal{O}$  (e.g. the $W$ transverse
mass) and the superscript refers to the pseudodata or to
the $j-$th template. The value of $\mw$ used in the template which
minimizes $\chi^2_j$ is considered as the preferred value of $\mw$ and
the difference $\Delta \mw=\mw -\mw^0$,  
is the
shift induced by the PDF set chosen for that set of pseudodata. 
A  similar approach has been used in~\cite{CMNTW,CMNTZ,atlasnote}.

\begin{figure}[t]
\begin{center}
\includegraphics[width=0.65\textwidth]{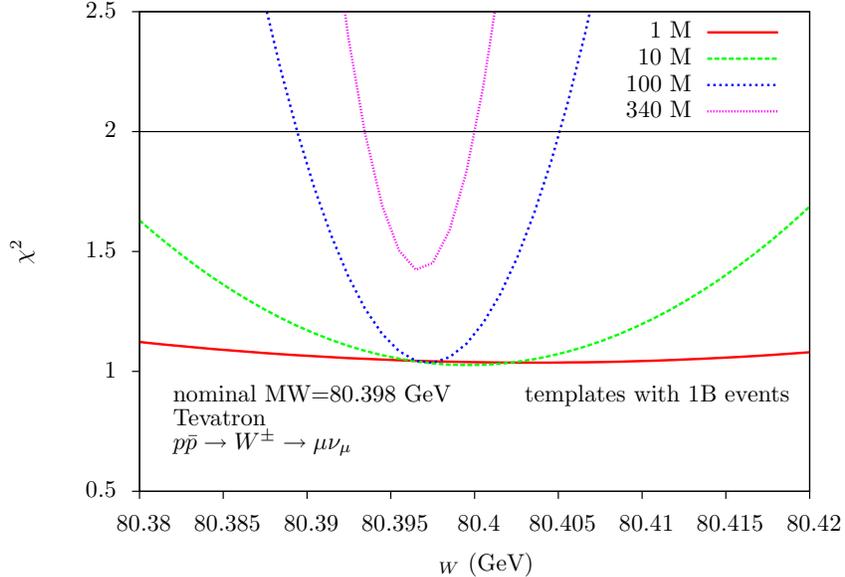}~
\end{center}
\caption{\small $\chi^2$ distributions obtained fitting Born level
  pseudodata with Born level templates for the same fixed PDF set at the
  Tevatron kinematics. The different curves correspond to different
  pseudodata samples each with different statistics. The
  $\Delta\chi^2=1$ rule indicates the resolution, at 68\% C.L., on the
  $W$ mass.}
\label{chi2valid}
\end{figure}

The fitting procedure has been validated by using samples of pseudodata
that have been produced with the same inputs and the same event generator
of the templates but with different statistics. 
In this case the function $\chi^2$ defined in Eq.~\ref{chi2j} can be
used to make a $\chi^2-$test.
When fitting pseudodata obtained with a given event generator, the nominal value of
$\mw$ used in the generation of the data is rediscovered as preferred
value within an interval determined  by the condition $\Delta\chi^2=1$,
which can be interpreted as a 68\% C.L. interval. This interval shrinks
as the number of events considered increases and correspondingly their
statistical fluctuations are damped, as shown in Fig.~\ref{chi2valid}.
In this example the templates have been generated
with 1B events while the pseudodata has been generated
with increasing statistics from 1M to 340M events. We
also checked that, fitting 1000 independent samples of Born
level pseudodata, the corresponding minima follow the $\chi^2$
distribution, as expected.

When the statistics of the pseudo-data become close
to those of the templates, the $\chi^2$ can deteriorate
since it becomes sensitive to the statistical fluctuations
of the latter, not accounted for in Eq.~\ref{chi2j}. This imposes a
practical limit on how accurate the pseudodata can be. This effect can
be seen in Fig.~\ref{chi2valid} for the case of pseudo-data generated
with 340M events. We find that a good compromise between resolution and
stability with respect to fluctuation is provided by using templates of 1B
events with pseudo-data generated with 100M events.

\subsection{Event generation}
\label{icbc}
Let us discuss now how the theoretical predictions of the DY kinematic
distributions have been generated.
We have studied the production process $p\bar p \to \mu^+ + X$ at the
Tevatron Run II ($\sqrt{s}$ = 1.96~TeV). We also consider the two
processes $pp \to \mu^+ + X$ and $pp \to
\mu^- + X$ at the LHC for $\sqrt{s}$ = 7 TeV and $\sqrt{s}$ = 14~TeV
center--of--mass energies. In the absence of QED effects, not considered
here, our results will be identical to those obtained with electrons
instead of muons. 
The numerical results have been obtained using  the following values for
the input parameters:
\begin{center}
\begin{tabular}{lll}
$G_{\mu} = 1.16637~10^{-5}$ GeV$^{-2}$ & 
$\mw = 80.398$~GeV&
$\mz=91.1876$~GeV \\
$\gw = 2.141$~GeV & 
$\sdw = 1 - \mw^2/\mz^2$&
$\mh = 120$~GeV\\
$V_{cd}=0.222$ &
$V_{cs}= 0.975$ &
$V_{cb}=0$ \\
$V_{ud}=0.975$ &
$V_{us}=0.222$ &
$V_{ub}=0$ \\
$V_{td}=0$ &
$V_{ts}=0$ &
$V_{tb}=1$ \\
\end{tabular}
\end{center}
The charm quark  in the partonic cross section is treated as a massless
particle, while the bottom quark does not contribute because of the
vanishing top density in the proton.

In the generation of Drell-Yan charged
current events we used the selection criteria summarized in 
Table~\ref{tab:selection}. These kinematic cuts are
similar to those used in the
corresponding experimental
analysis. Note that the main difference between the Tevatron and
LHC cuts is a wider acceptance for the rapidity of the leptons
in the latter case.
The $W$ transverse mass distribution has been studied in the interval
$50$ GeV$\leq M_\perp^W\leq100$ GeV, with a bin size of 0.5 GeV, since the
jacobian peak region is the most sensitive for the determination
of $\mw$.
All the following analysis are performed with bare leptons both in the
pseudodata and in the templates. 

\begin{table}[t]
\begin{center}
\begin{tabular}{|c|c|}
\hline
Tevatron & LHC \\
\hline
$p_{\perp}^{\mu} \geq$~25 GeV        & $p_{\perp}^{\mu} \geq$~25 GeV \\
$\rlap{\slash}{\! E_T} \geq$~25 GeV &  $\rlap{\slash}{\! E_T} \geq$~25
GeV \\ 
$|\eta_\mu|< 1.0$                   & $|\eta_\mu|< 2.5$ \\
\hline
\end{tabular}
\caption{\small Selection criteria for $W^{\pm}\to l^{\pm}\nu$ events
for the Tevatron and the LHC.}
\label{tab:selection}
\end{center}
\end{table}

The pseudodata and the templates have been generated using the following
event generators: at Born level with HORACE~\cite{CMNV},
at NLO-QCD with DYNNLO~\cite{dynnlo} and at NLO+NNLL-QCD with
ResBos~\cite{resbos}. These generators  allow to compute the
distributions of the final state leptons in the DY processes at various
perturbative orders.
For example ResBos includes, on top of the NLO-QCD corrections, part of
the NNLO-QCD terms matched with the resummation of the large $\log\left(
\frac{p_\perp^W}{\mw}\right)$ at leading logarithmic (LL) and
next-to-leading logarithmic (NLL) accuracy, and has been widely used at the
Tevatron. 

Our final results for the determination of $\mw$ will be those
obtained at NLO-QCD with DYNNLO, although, as we will show below,
the qualitative results are already very similar at Born level.

\subsection{PDF uncertainties}
\label{densities}

The proton PDF sets considered in this study are the three
global sets that include all the relevant hard scattering
 data. In particular we will use 
the NLO-QCD CTEQ6.6~\cite{Nadolsky:2008zw}, 
MSTW2008~\cite{Martin:2009iq} and NNPDF2.1~\cite{Ball:2011mu} 
PDF sets. Each collaboration provides a prescription to
estimate the PDF uncertainties: in particular we recall the formula for the symmetric error in the Hessian approach (CTEQ,MSTW)
\be
\Delta X = \frac12 \sqrt{\sum_{i=1}^N\left[X^+_i-X^-_i \right]^2  }
\ee
and the average over the ensemble of PDF replicas (NNPDF)
\be
\langle\mathcal{F}
[\{q\}]\rangle=\frac{1}{N_{rep}}\sum_{k=1}^{N_{rep}}\mathcal{F}[\{q^{(k)}\}] \nonumber
\ee
\be
\sigma_{\mathcal{F}}=\left(\frac{1}{N_{rep}-1}
\sum_{k=1}^{N_{rep}}\left(\mathcal{F}[\{q^{(k)}\}]-
\langle\mathcal{F}[\{q\}]\rangle\right)^2\right)^{1/2} . \nonumber
\ee
We refer to the original publications as well as to the recent
reviews~\cite{Forte:2010dt,DeRoeck:2011na,Alekhin:2011sk} for more details.
Let us recall that the use of the three global PDF sets is the
basis of the current PDF4LHC recommendation~\cite{Botje:2011sn} 
for the use of PDFs
in the analysis of LHC data.

On top of the PDF uncertainties that arise from the experimental
uncertainties of the data used in their determination, there
are other sources of theoretical uncertainties closely related
to PDFs. In the first place,
PDFs are correlated with the value of the strong coupling
constant $\alpha_s\lp m_Z\rp$ used in the PDF determination,
expecially the gluon PDF. Again, all three
groups provide prescriptions on how to combine
the PDF and strong coupling uncertainties in a consistent way.
A summary of the prescriptions recommended by each group can
be found in the
PDF4LHC working group interim report~\cite{Alekhin:2011sk} (see also
Ref.~\cite{LHCHiggsCrossSectionWorkingGroup:2011ti}). A practical guide
on the  way to efficiently implement the recommendations by the
different groups can be found in Ref.~\cite{practicalguide}. 
While the impact of variations on the value of $\alpha_s\lp M_Z\rp$ are
known to be small for vector boson production\footnote{As opposed to other
relevant LHC processes, like Higgs boson production via gluon fusion,
where $\alpha_s$ uncertainties can be the dominant theoretical 
uncertainty~\cite{Demartin:2010er}.}, they may need to be taken into
account at the level of precision required for the determination of
$\mw$. 

On top of the value of the strong coupling, PDFs depend 
as well on the value of the heavy quark masses
$m_c$ and $m_b$ due to two different reasons. The first one is
the fact that even though most LHC perturbative computation are done up 
to power-suppressed terms, terms
of $\mathcal{O}\lp m_h^2/Q^2\rp$ do still have a 
non-negligible impact on PDF fits, expecially to the HERA collider data.
Power suppressed terms are accounted for in the various General--Mass
VFN schemes used in modern PDF
sets~\cite{Thorne:2006qt,Forte:2010ta,Aivazis:1993pi}, and the choice of
$m_c$ affects the GM-VFN predictions and thus the fitted PDFs.
Different GM-VFN schemes have been compared in the Les Houches
heavy quark benchmark study \cite{Binoth:2010ra}, elucidating their
differences and similarities. 

The second reason has simply to do with the fact that 
heavy quark PDFs are obtained
by assuming them to vanish at threshold, and then to be generated by 
perturbative evolution. But changing the mass also changes the position
of the threshold, and thus the heavy quark PDFs (and their contribution
to the cross section) depend on the value of $m_h$. For example, for $W$
production, the initial state with one charm and one
strange quarks occurs at the Born level approximately in the 7\% of the 
cases at the Tevatron, in the 16\% at LHC 7 TeV for $W^+$ production and
in the 25\% for $W^-$ production, in the 24\% at LHC 14 TeV for $W^+$
production and in the 32\% for $W^-$ production.  

For these reasons the precise value of the $m_c$ has an impact on  the 
kinematic distributions from which $\mw$ is extracted and must be
accounted for, expecially since charm mass variations are known to induce sizeable effects for $W$ production at colliders~\cite{Martin:2010db,Ball:2011mu}.

\section{PDF uncertainties for the transverse mass distribution}
\label{resultsmtw}

Now that the setup of the analysis has been presented, we consider how
PDFs and related uncertainties affect the lepton pair kinematic
distributions, in particular the transverse mass distribution, and in
the next section we will consider their impact on the determination of
$\mw$. 
As discussed in Sect.~\ref{fitting}, only those sources of uncertainties
that induce distortions on the shape of the distribution (rather than
on its normalization) will have an impact for the extraction of $\mw$.

The transverse mass distribution has the advantage, with respect to
the lepton $p_T$ distribution, that QCD-NLO corrections are rather
moderate and in particular have a small effect on the shape
of the distribution. Experimental issues in its measurement
like the systematic uncertainties due to the neutrino $p_T$
reconstruction are not addressed here. 

In the following we will consider two related distributions: the
transverse mass distribution,
\be
\label{eq:trans_mass_dist}
\mathcal{O}\lp M_{\perp}^W \rp\equiv\frac{d\sigma}{dM_{\perp}^W}\lp
M_{\perp}^W \rp, \qquad M_{\perp}^{W}=\sqrt{2p_t^lp_t^{\nu}\left(
    1-\cos\lp \phi^l-\phi^{\nu}\rp\right)} \ ,
\ee
and the same distribution but normalized to the integrated cross section
in the region used for the $\mw$ fit,
\be
\label{eq:trans_mass_dist_norm}
\widetilde{\mathcal{O}}\lp M_{\perp}^{W}\rp\equiv\frac{1}{\sigma^{\rm fit}}
\frac{d\sigma}{dM_{\perp}^W}\lp M_{\perp}^{W}\rp, \qquad \sigma^{\rm fit}
\equiv \int_{M_{\perp}^{W,\rm min}}^{M_{\perp}^{W,\rm
    max}}dM\frac{d\sigma}{dM_{\perp}^W}\lp M\rp \, ,
\ee
with $M_{\perp}^{W,\rm min}=50$ GeV and $M_{\perp}^{W,\rm min}=100$
GeV. The motivation to define ${\tilde {\cal O}}$ is that
in this way normalization effects, irrelevant for the $\mw$ determination,
cancel out, and one is left only with the contribution of
PDF uncertainties that induce shape distorting effects.
The use of normalized distributions has also been adopted
in the Tevatron analysis~\cite{:2009nu}.

\begin{figure}[t]
\begin{center}
\includegraphics[height=55mm]{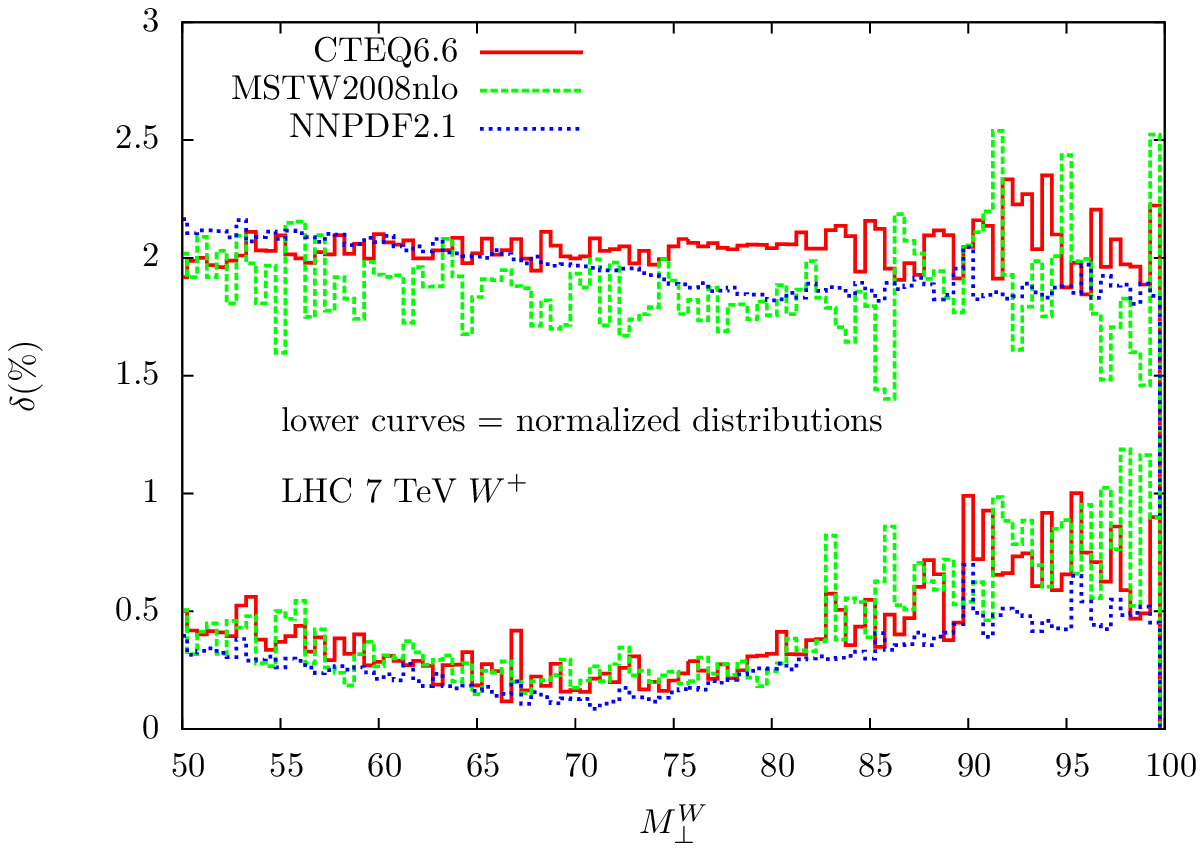}
\includegraphics[height=55mm]{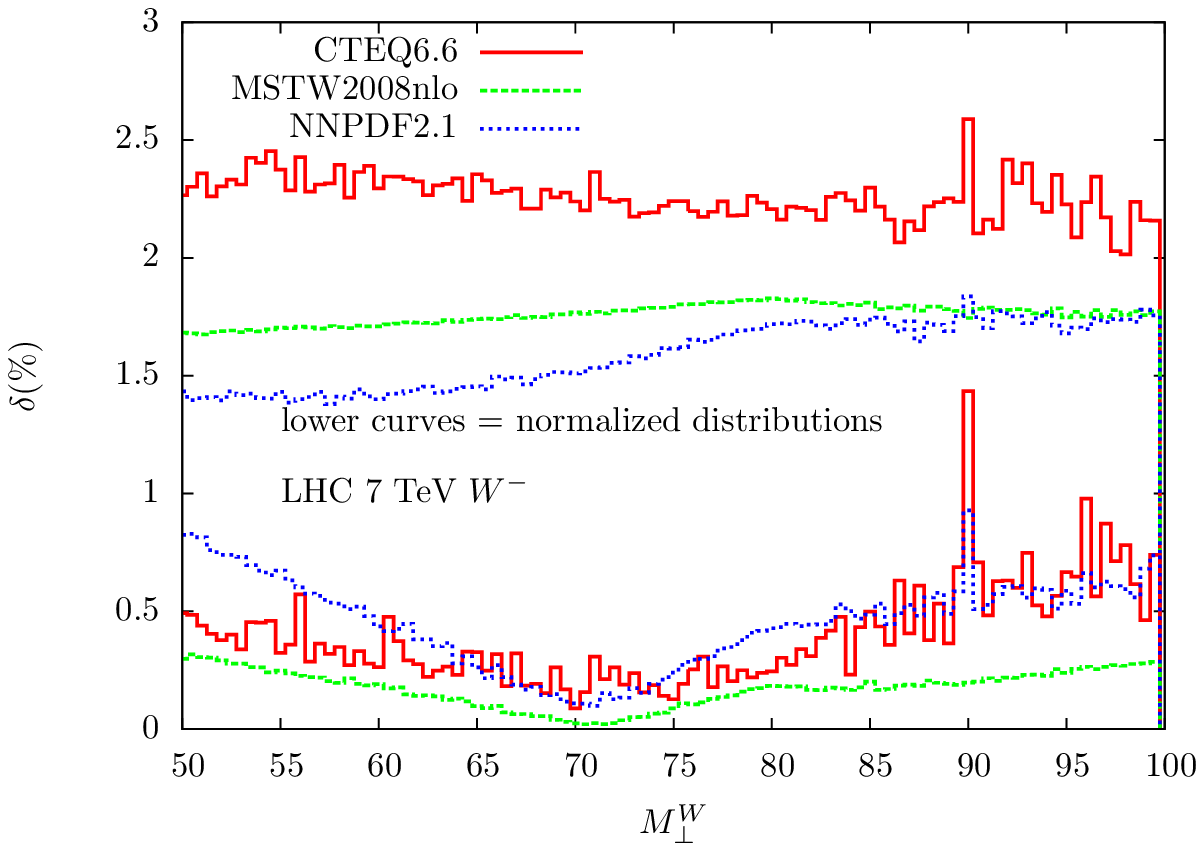}\\
\includegraphics[height=55mm]{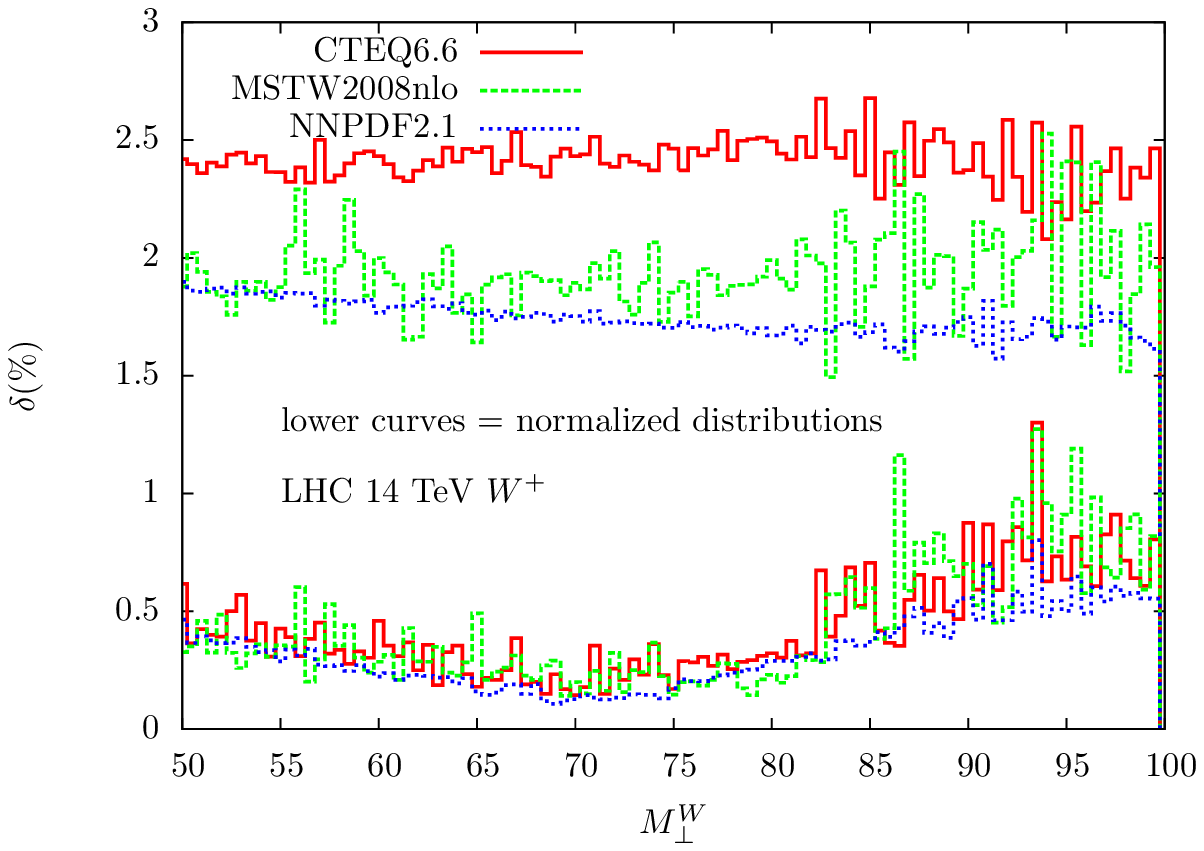}
\includegraphics[height=55mm]{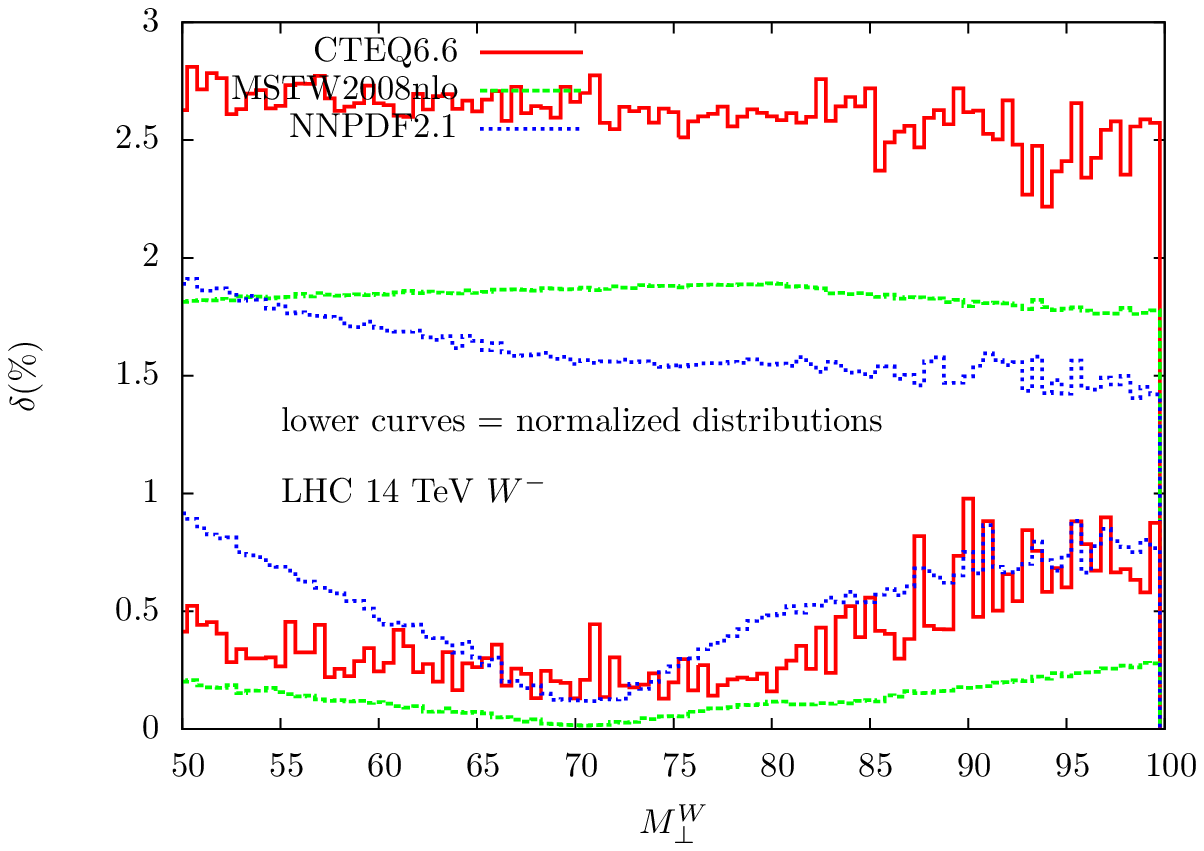}
\end{center}
\caption{\small Relative PDF uncertainties in the Born level
transverse mass distributions, computed with respect
the respective central
PDF set. From top to bottom: LHC 7 TeV $W^+$ and $W^-$ and LHC 14 TeV
$W^+$ and $W^-$. Both 
the PDF uncertainties on the standard distribution,
Eq.~\ref{eq:trans_mass_dist}, and the normalized distribution,
Eq.~\ref{eq:trans_mass_dist_norm}, are shown.} 
\label{fig:mtw-uncBorn}
\end{figure}

\begin{figure}[t]
\begin{center}
\includegraphics[height=55mm]{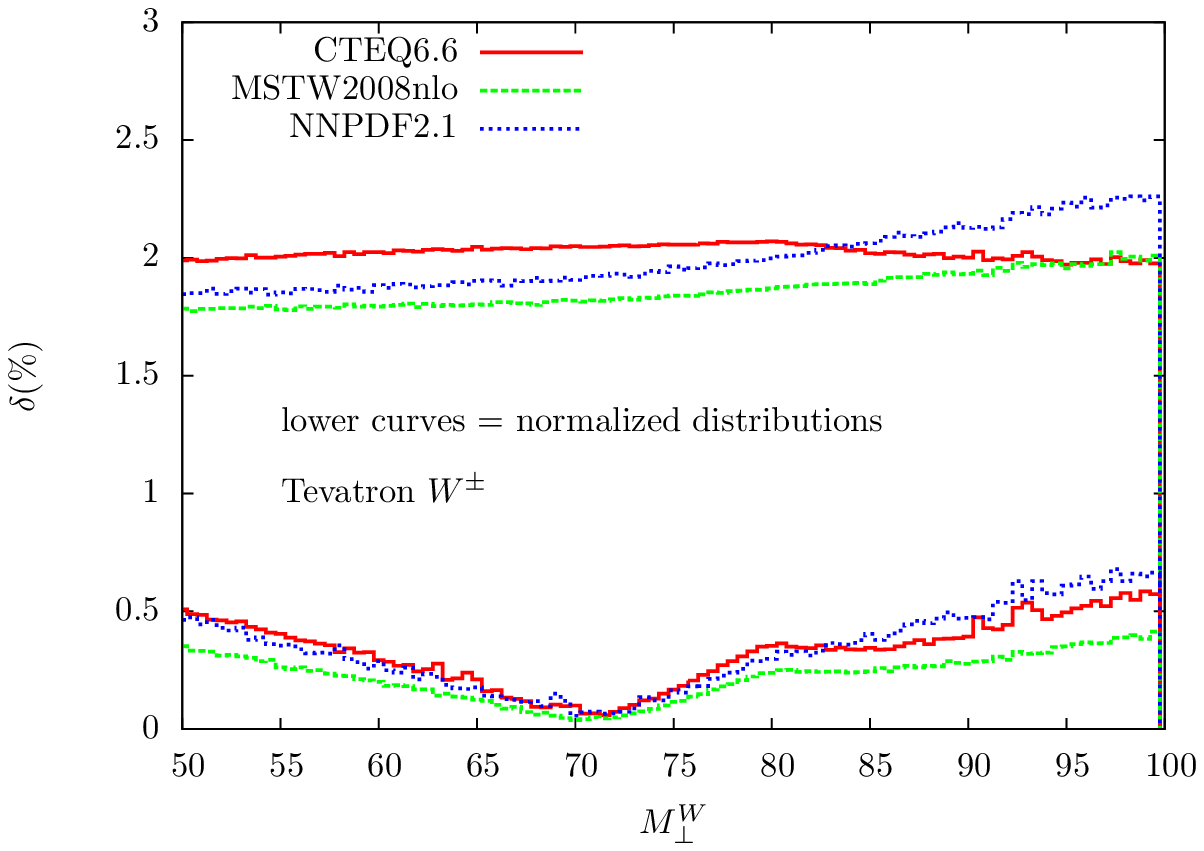}\\
\end{center}
\caption{\small Same as Fig.~\ref{fig:mtw-uncBorn} for the Tevatron.} 
\label{fig:mtw-uncBorn-TEV}
\end{figure}

The same NLO PDFs are used both to generate the Born and NLO-QCD
distributions. 
In Figs.~\ref{fig:mtw-uncBorn} and~\ref{fig:mtw-uncBorn-TEV} we
compare, for the three PDF sets, the relative size of the
pure PDF uncertainties, at the Tevatron and at the LHC 7 and 14 TeV,
for the transverse mass distributions computed at Born level
with the HORACE generator.
In the latter case we consider separately the two cases of $W^+$ and
$W^-$ production, since in a proton-proton collider the two
distributions are different unlike in a proton-antiproton collider. We
show both the standard, Eq.~\ref{eq:trans_mass_dist}, and the
normalized, Eq.~\ref{eq:trans_mass_dist_norm}, distributions.

We observe that the PDF uncertainties in the normalized distributions
are much smaller than in the standard transverse mass
distributions: the reason for this is that
variations in the normalization of the distribution, which are not
relevant for the determination of $\mw$, cancel out in the normalized
distributions. 
Note that from Figs.~\ref{fig:mtw-uncBorn} and~\ref{fig:mtw-uncBorn-TEV} we
see that PDF uncertainties are at the few permille level.

\begin{figure}[t]
\begin{center}
\includegraphics[height=55mm]{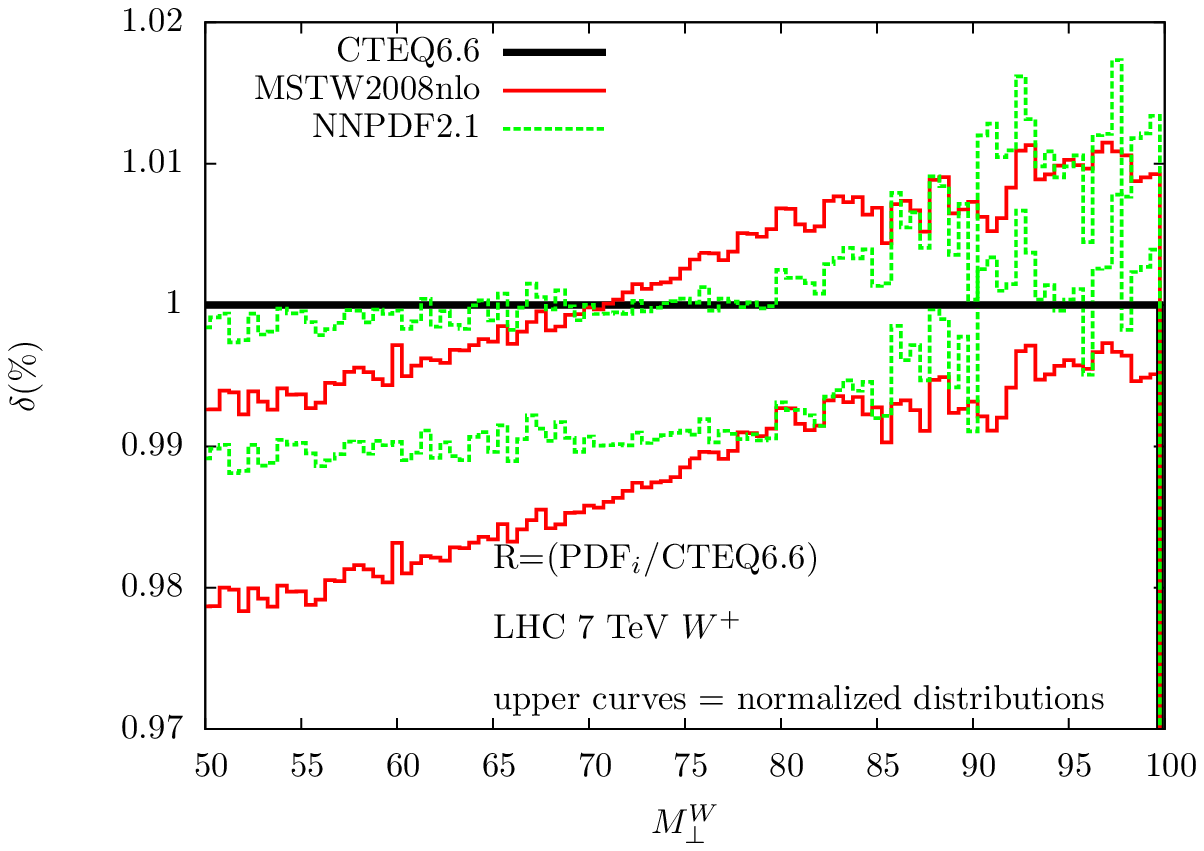}
\includegraphics[height=55mm]{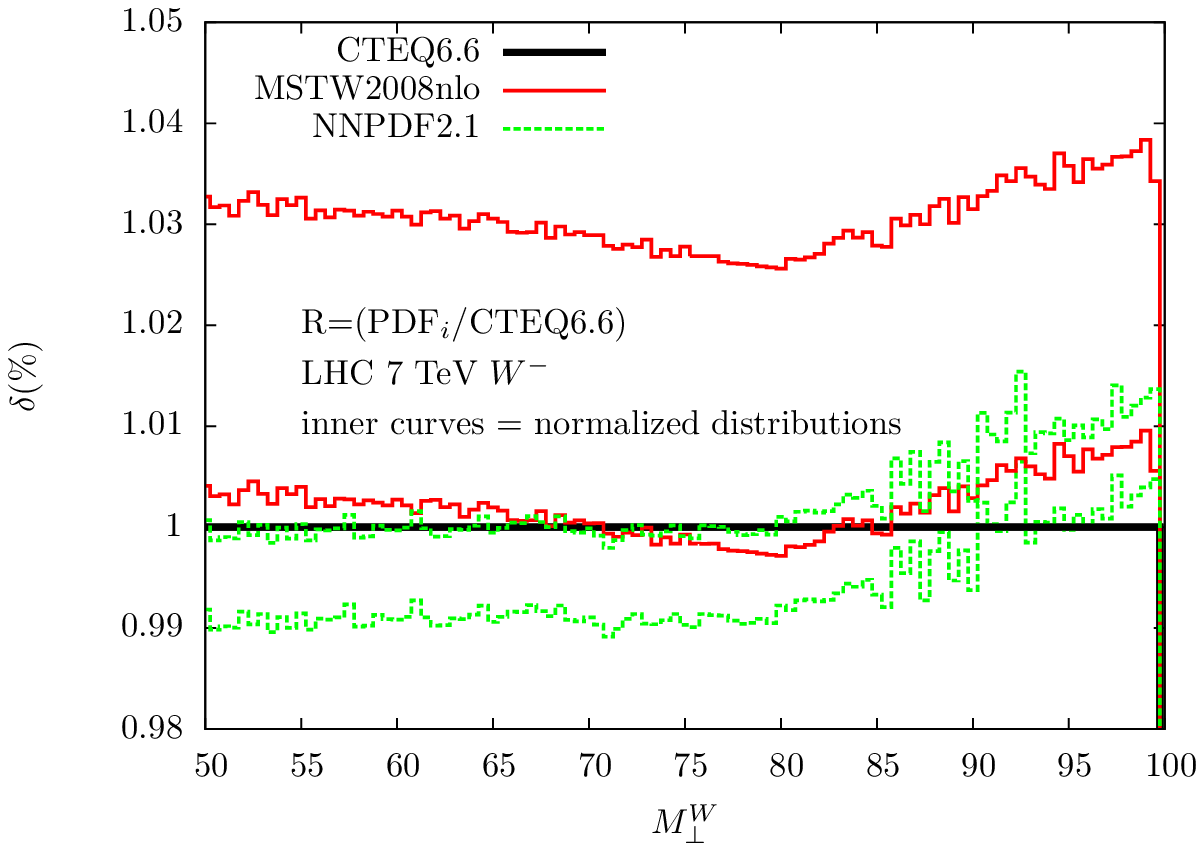}\\
\includegraphics[height=55mm]{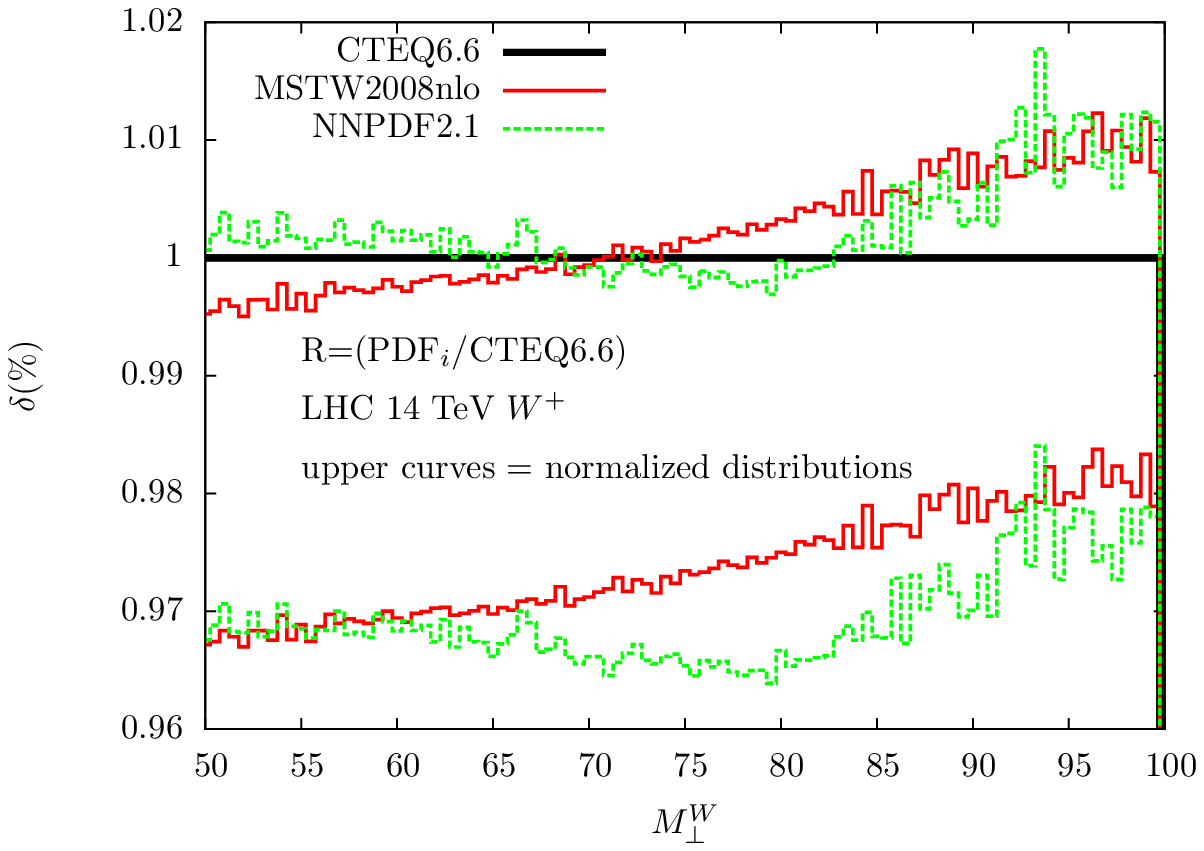}
\includegraphics[height=55mm]{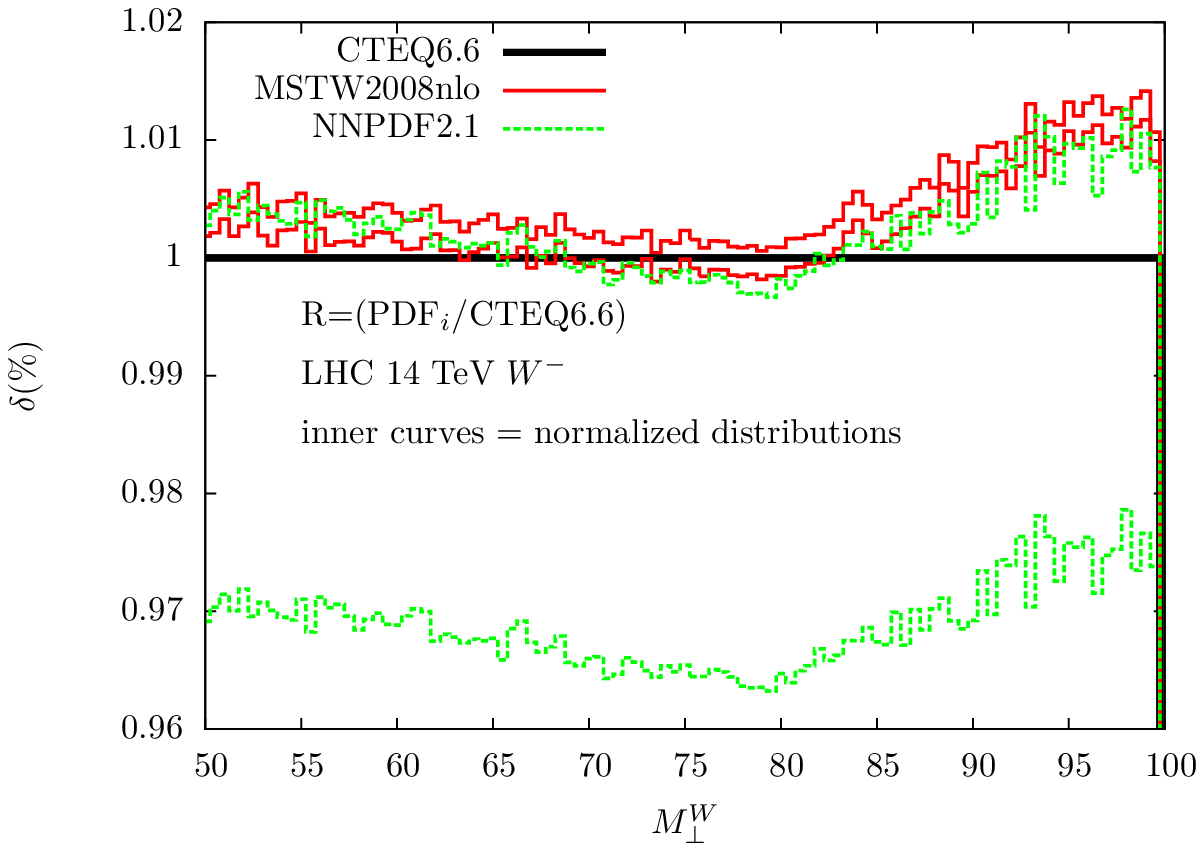}
\end{center}
\caption{\small Relative difference between the distributions
obtained with the central PDF set of CTEQ6.6, MSTW08 and NNPDF2.1,
normalized to the CTEQ6.6 result. We show the results both for
the normalized and for the standard distributions for LHC
7 TeV (upper plots) and 14 TeV (lower plots).
\label{fig:mtw-uncBorn-CTEQnorm} }
\end{figure}

\begin{figure}[t]
\begin{center}
\includegraphics[height=55mm]{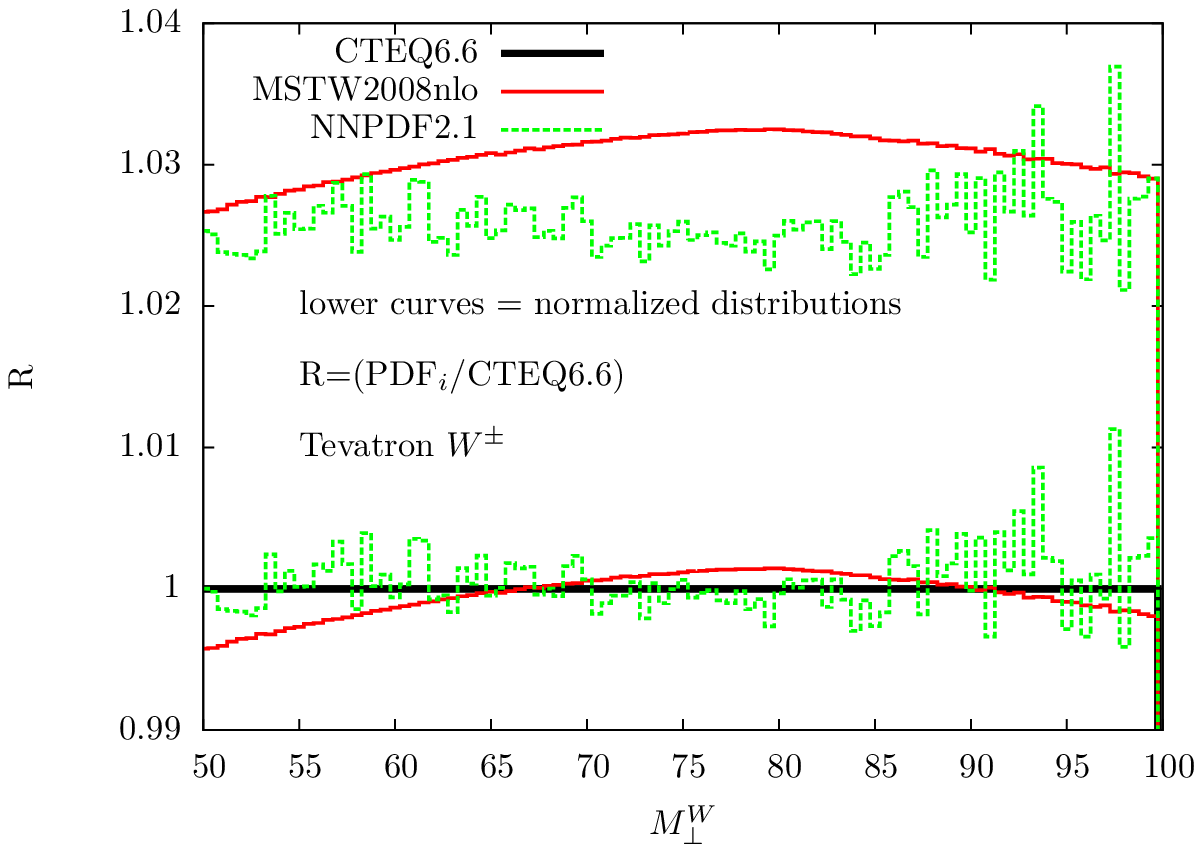}\\
\end{center}
\caption{\small Same as Fig.~\ref{fig:mtw-uncBorn-CTEQnorm} for the
  Tevatron.} 
\label{fig:mtw-uncBorn-CTEQnorm-TEV}
\end{figure}

The previous plots show that PDF uncertainties are similar
for the three global PDF sets. However, it could still be the case
that the distributions obtained with the central set of each
PDF set differ sizeably among them, leading to an uncertainty
in $\mw$ much larger than the nominal PDF uncertainty of a single
set. To check that this is not the case, in 
Figs.~\ref{fig:mtw-uncBorn-CTEQnorm} (for the LHC)
and~\ref{fig:mtw-uncBorn-CTEQnorm-TEV} (for the Tevatron) we show 
the ratio of transverse mass distributions for each central PDF
set normalized to the central CTEQ6.6 predictions.

The results of Figs.~\ref{fig:mtw-uncBorn-CTEQnorm}
and~\ref{fig:mtw-uncBorn-CTEQnorm-TEV}  show that, while
the standard transverse mass distributions differ at the few percent level
between different PDF sets, the normalized distributions on the other
hand are much more similar, providing an excellent
agreement of the central values and differing only at the permille level. 
This is the same order of magnitude as the intrinsic PDF
uncertainties. This suggests that the determinations
of $\mw$ from the three different sets are consistent
within the respective PDF uncertainties: we will explicitly
verify this expectation in Sect.~\ref{resultsmw}.

\begin{figure}[h]
\begin{center}
\includegraphics[height=55mm]{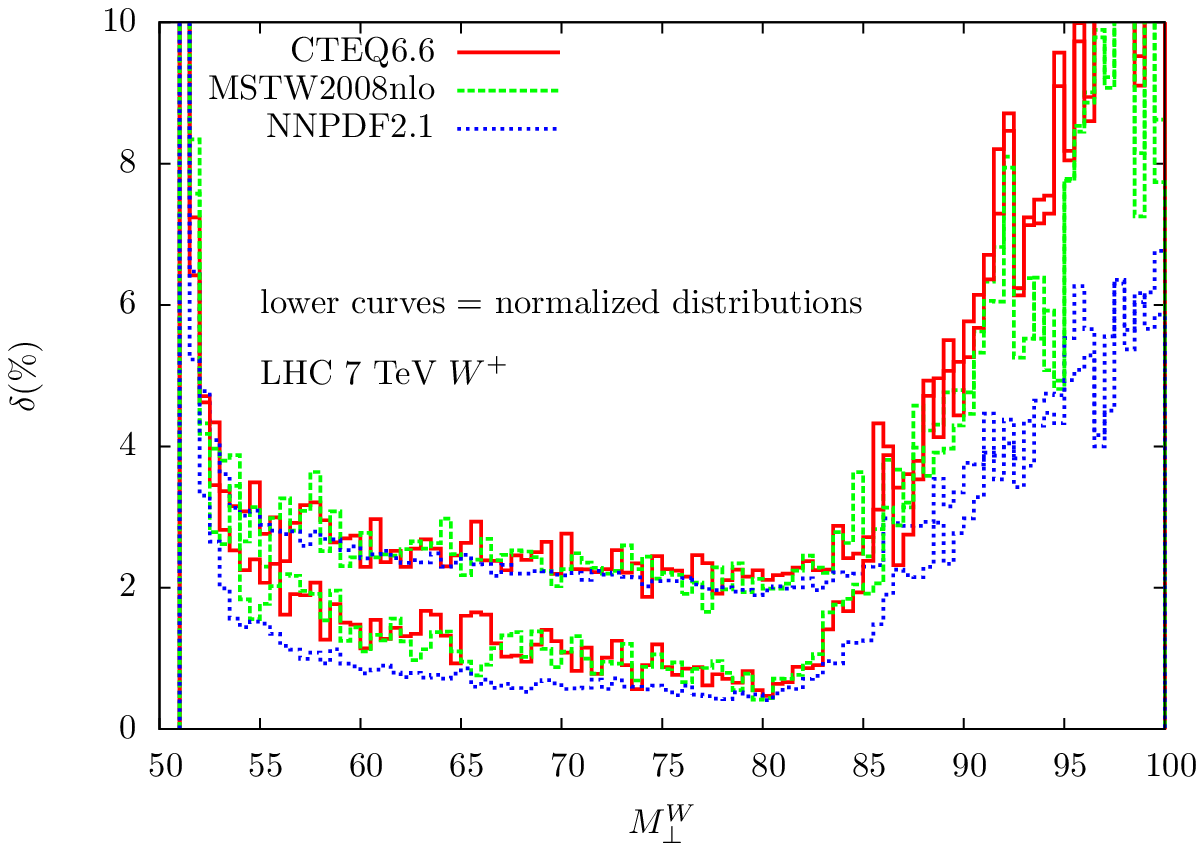}
\includegraphics[height=55mm]{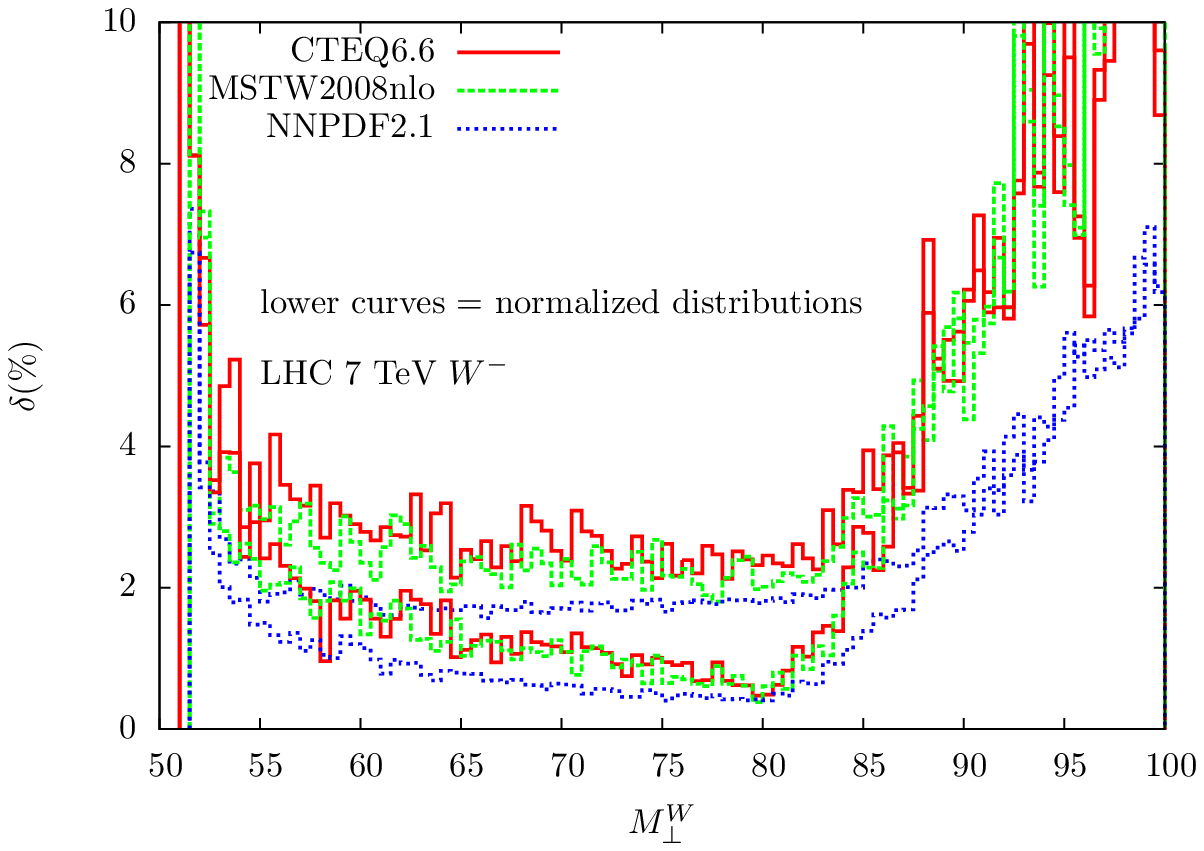}\\
\includegraphics[height=55mm]{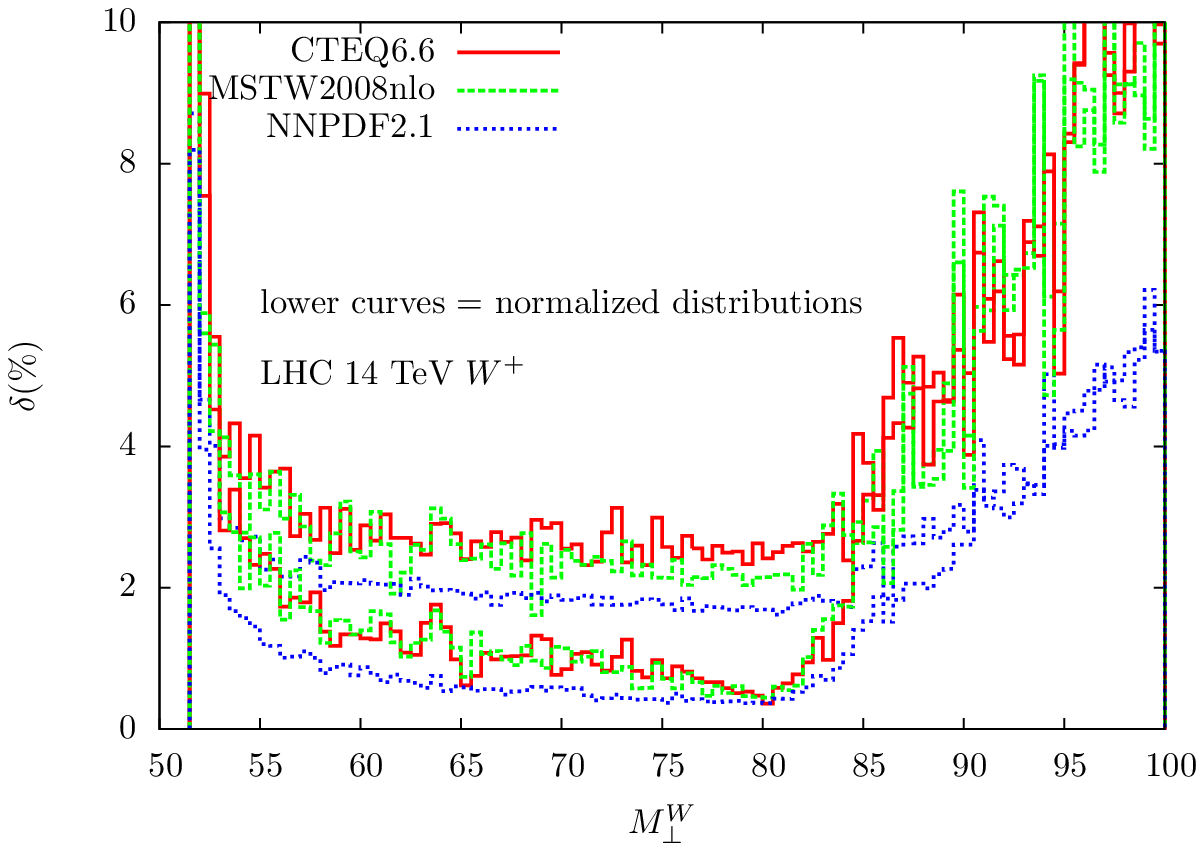}
\includegraphics[height=55mm]{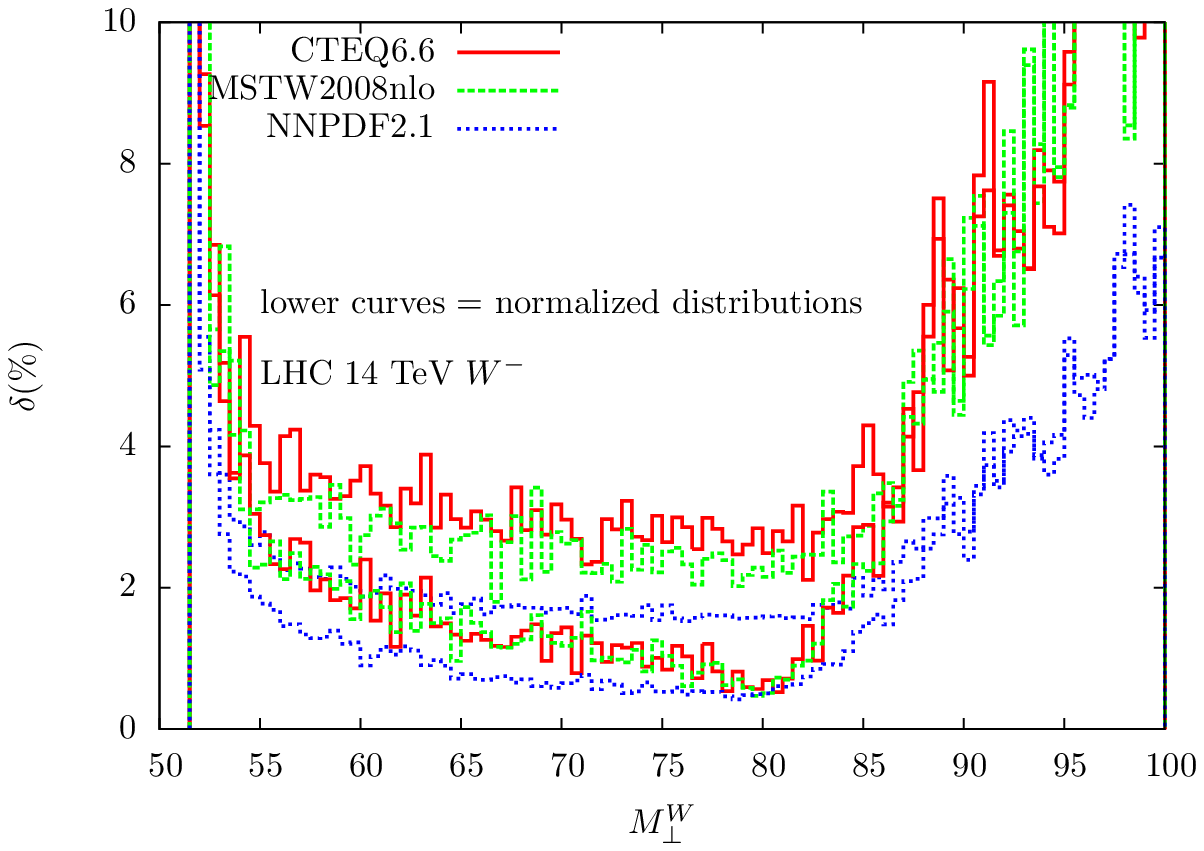}
\end{center}
\caption{\small Same as Fig.~\ref{fig:mtw-uncBorn} but now
the transverse mass distributions have been computed at NLO-QCD
using the event generator DYNNLO.} 
\label{fig:mtw-uncQCD}
\end{figure}

\begin{figure}[h]
\begin{center}
\includegraphics[height=55mm]{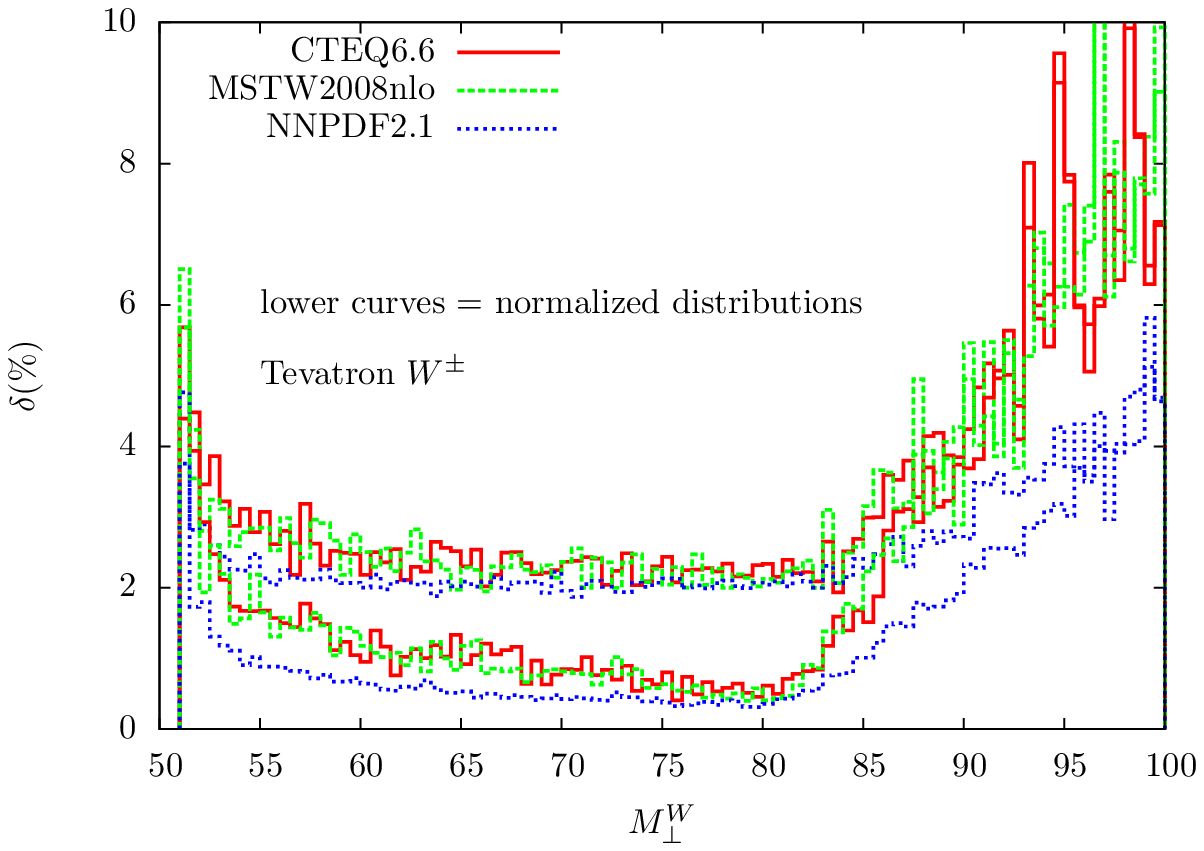}\\
\end{center}
\caption{\small Same as Fig.~\ref{fig:mtw-uncBorn-TEV} but now
the transverse mass distributions have been computed at NLO-QCD
using the event generator DYNNLO.} 
\label{fig:mtw-uncQCD-TEV}
\end{figure}

The uncertainties on the transverse mass distributions, this
time computed with the DYNNLO generator at NLO-QCD, are shown in 
Figs.~\ref{fig:mtw-uncQCD}
 and~\ref{fig:mtw-uncQCD-TEV}.
The QCD corrections introduce a new partonic sub-process ($qg\to q l
\nu_l$) and the related gluon density uncertainty.
The latter induces an increase
of PDF uncertainties in the large tail of the transverse mass
distribution above the jacobian peak, where the cross section steeply
falls, as well as for small transverse masses. On the
other hand, in the region near the peak, most relevant for the
determination of $\mw$, the PDF uncertainties at NLO-QCD are similar to
those of the Born distributions.

In the transverse mass distributions normalized to their respective
cross sections, 
the difference in PDF normalization  has been removed and
the uncertainty is due only to the different shapes induced by the PDF
sets considered. The comparison in Figs.~\ref{fig:mtw-uncQCD}
 and~\ref{fig:mtw-uncQCD-TEV} shows that the typical size of the PDF
 uncertainty on these normalized observables is well below the 1\% level,
whereas in the non-normalized case it ranges between 2 and 3\%.
The latter are the typical PDF uncertainties that one finds for the
inclusive cross section~\cite{Ball:2011mu}.

\begin{figure}[t]
\begin{center}
\includegraphics[height=60mm]{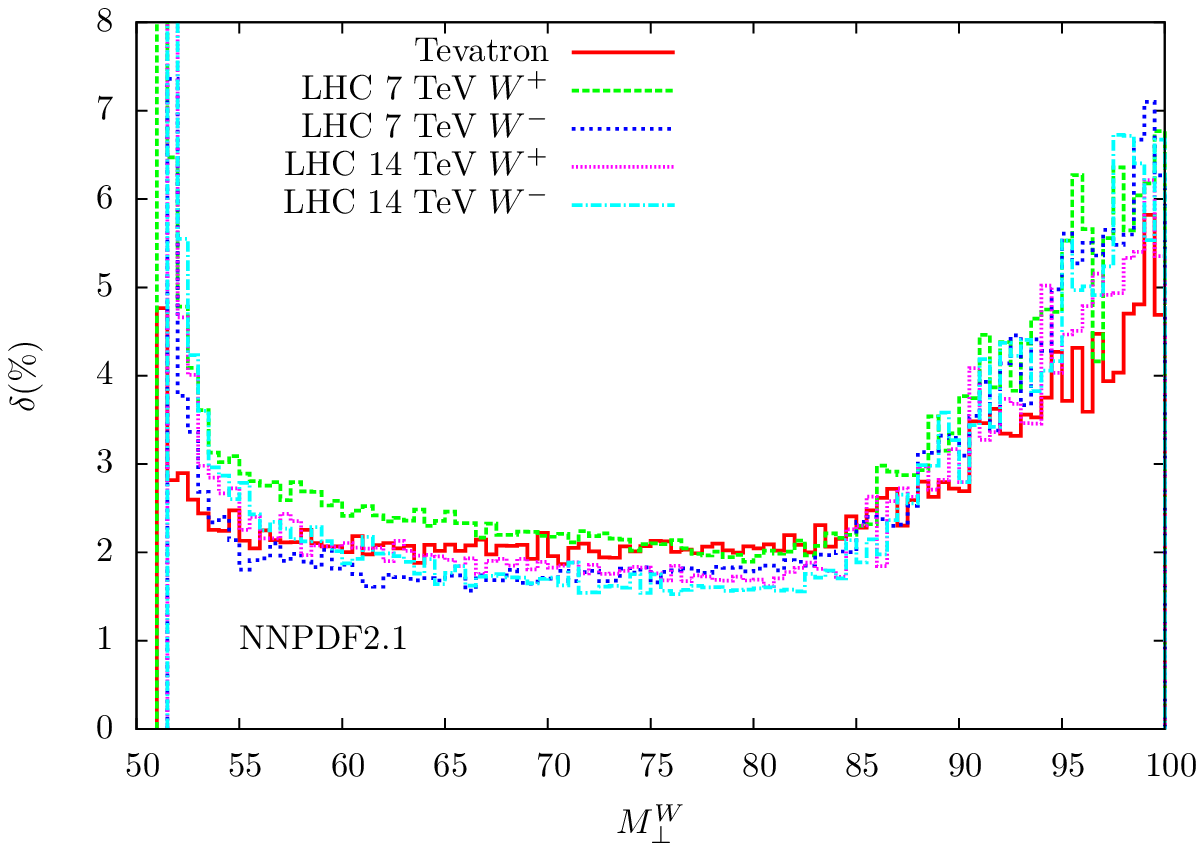}
\end{center}
\caption{\small The relative PDF uncertainty in the 
 standard transverse mass distributions 
for NNPDF2.1  for different colliders, energy, and final states.} 
\label{fig:mtw-uncNNPDF}
\end{figure}

In Fig.~\ref{fig:mtw-uncNNPDF} we show, in the case of the NNPDF2.1 set, 
how the PDF uncertainties in the NLO-QCD transverse mass
distribution varies with the energy, collider type and
final state. 
The different uncertainties are very similar in size
(e.g. they are all at 2\% level, below 80 GeV).
Fig.~\ref{fig:mtw-uncNNPDF} shows that PDF uncertainties
in the transverse mass distribution
are relatively independent of the collider and final state.
This result is reassuring  since it shows that, at least
from the PDF point of view, the determination of $\mw$ at the
LHC is not more challenging than at the Tevatron.

\begin{figure}[t]
\begin{center}
\includegraphics[height=60mm]{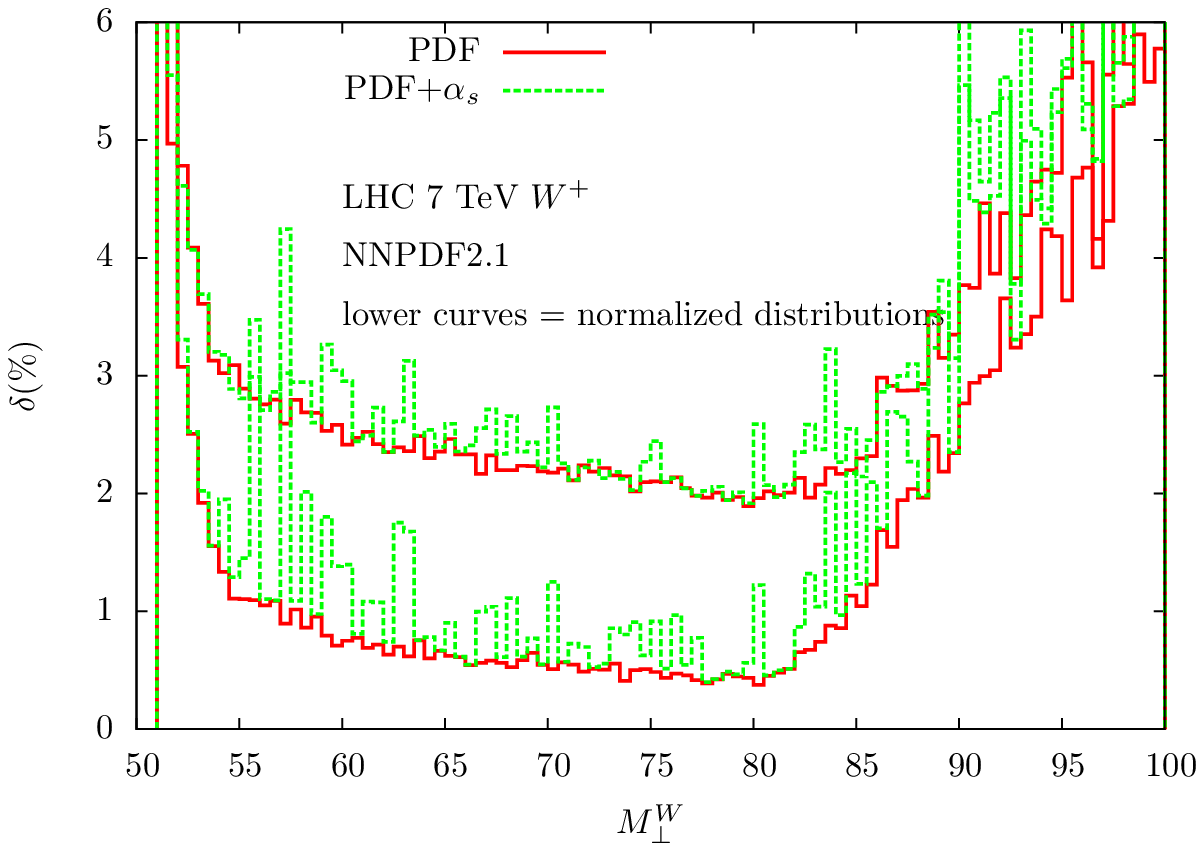}~
\includegraphics[height=60mm]{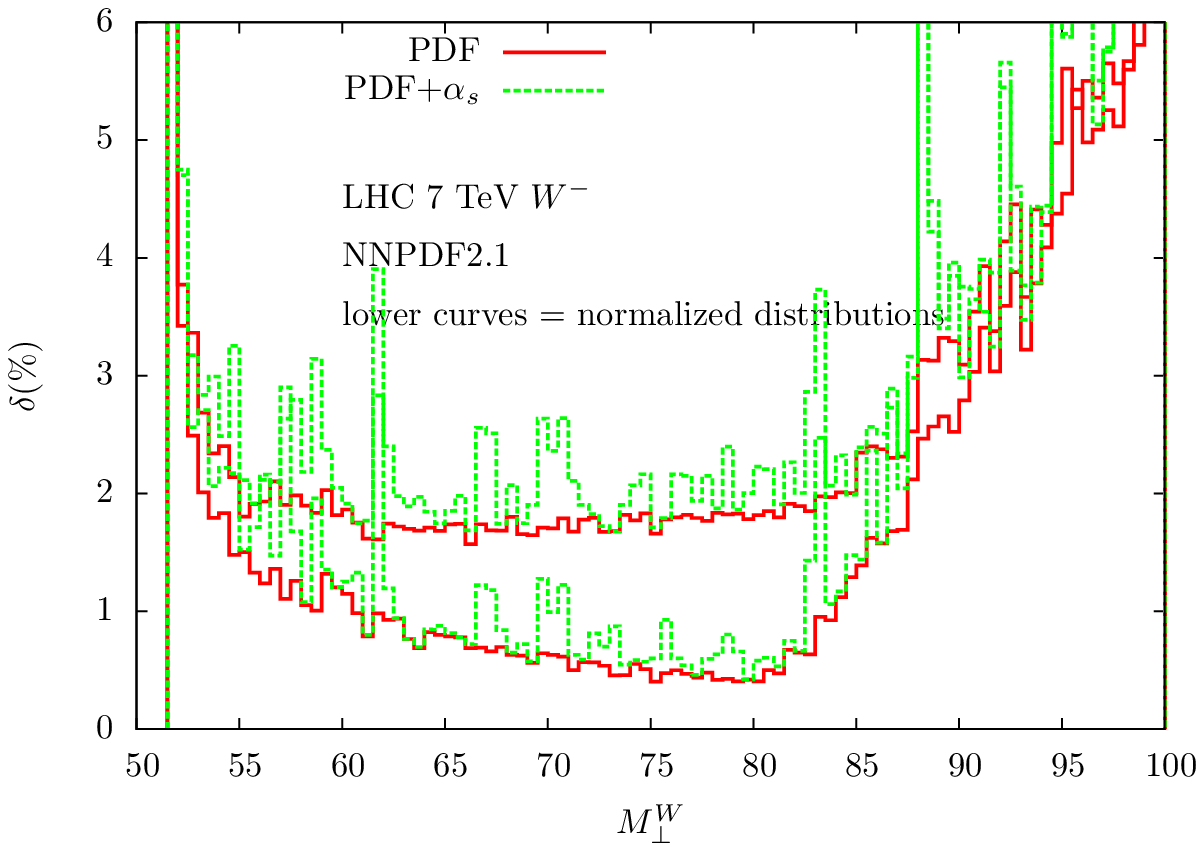}
\end{center}
\caption{\small Comparison of the
PDF--only uncertainty and the combined PDF+$\alpha_s$ uncertainty
 of the transverse mass distribution for NNPDF2.1.
For simplicity we show only the  distributions at the
LHC 7 TeV, the distributions for Tevatron and LHC 14 TeV
 are quantitatively very similar.
}
\label{fig:mtw-alphas}
\end{figure}

Let us now assess the impact of the uncertainties related 
to the values of $\alpha_s$ and $m_c$ on the transverse mass
distribution. In Fig.~\ref{fig:mtw-alphas} we show
for NNPDF2.1 the PDF--only uncertainty compared to the combined
PDF+$\alpha_s$ uncertainty. Following
Ref.~\cite{LHCHiggsCrossSectionWorkingGroup:2011ti} we assume that the
uncertainty on the strong couping is $\delta_{\alpha_s}=0.0012$ at the
68\% confidence level. For simplicity we show only the distributions at
the LHC 7 TeV: the distributions for Tevatron and LHC 14 TeV are
quantitatively very similar. We conclude that  $\alpha_s$ uncertainties
are negligigle as compared to the PDF uncertainties for this distribution.

\begin{figure}[t]
\begin{center}
\includegraphics[height=60mm]{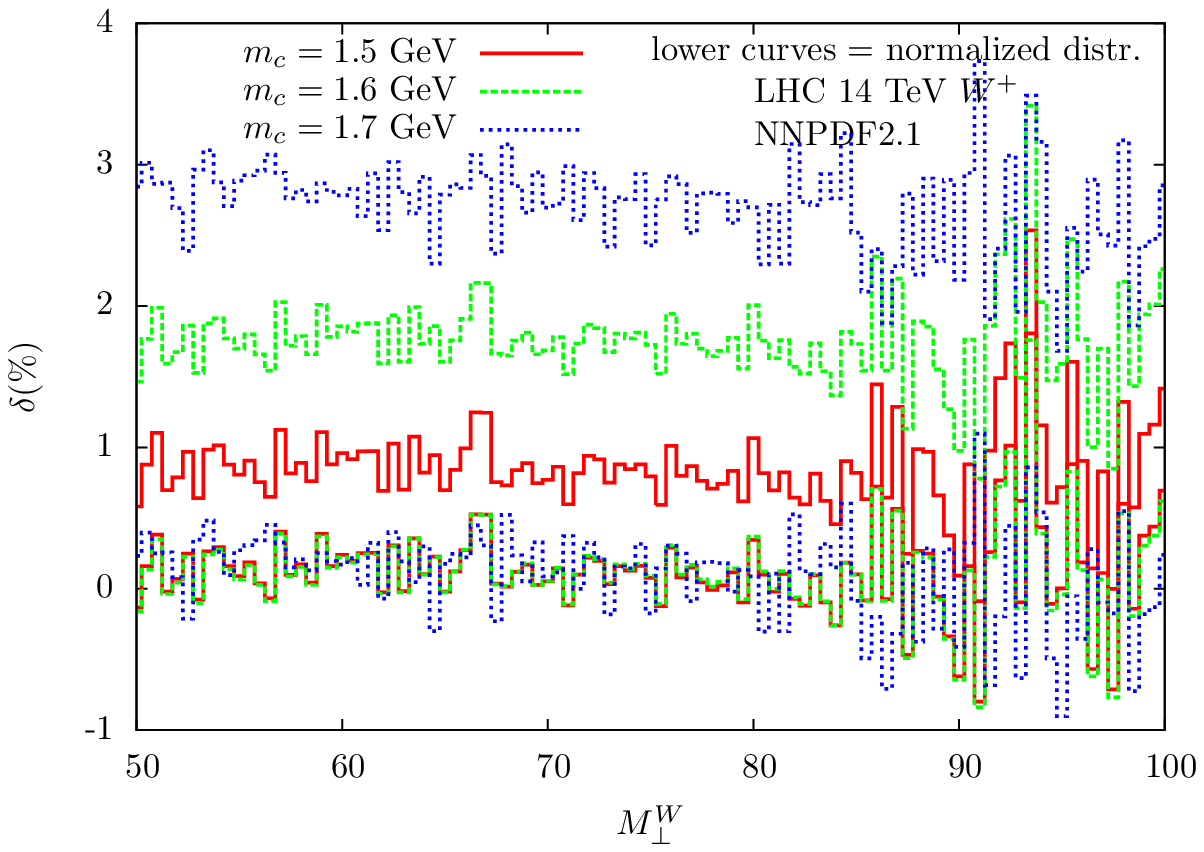}~
\includegraphics[height=60mm]{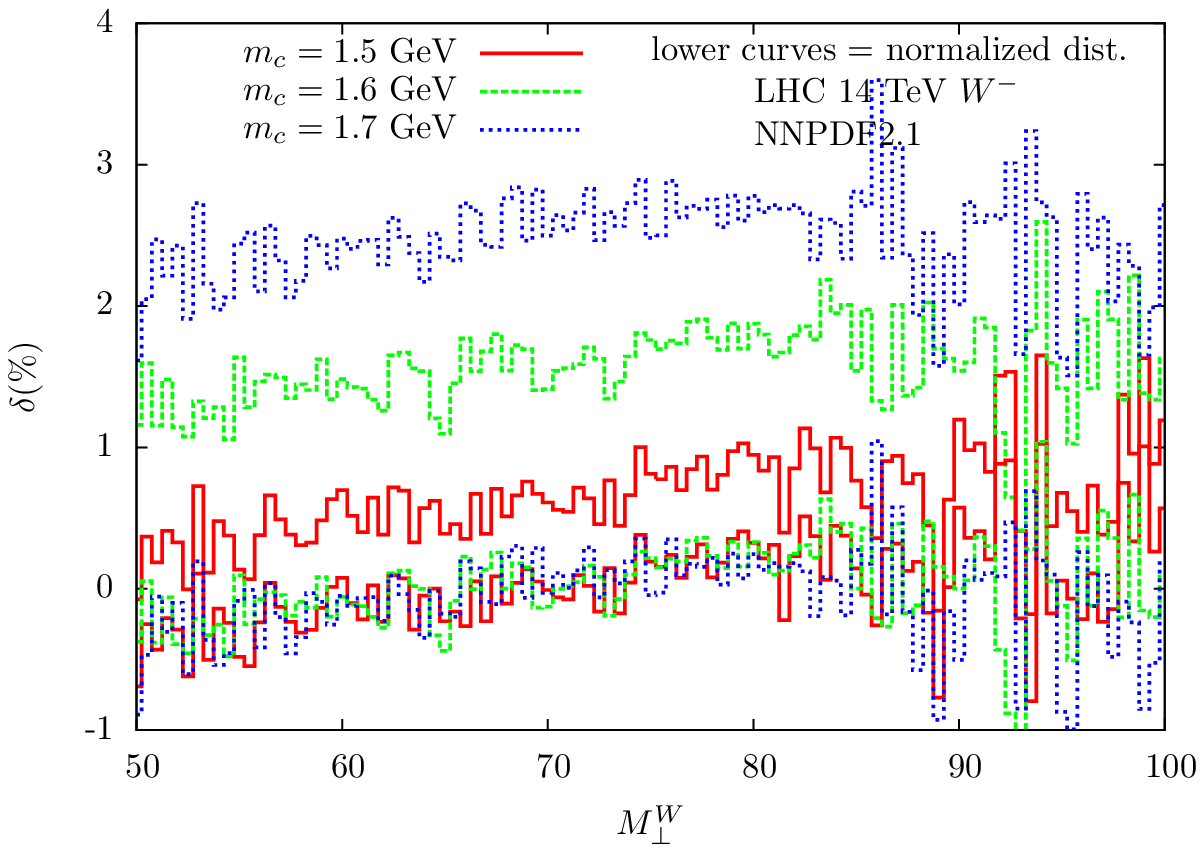}
\end{center}
\caption{\small For NNPDF2.1 we show the dependence on $m_c$
 of the transverse mass distribution, 
expressed as relative deviation from the central NNPDF2.1 set
with $m_c^2=$2 GeV$^2$. We show results both for the normalized and
for the standard transverse mass distributions. We consider only
the LHC 14 TeV case, where charm mass effects are known to be
more important.} 
\label{fig:mtw-charm}
\end{figure}

We have also studied the dependence of the results on the
value of $m_c$ used in the PDF determination (using the NNPDF2.1 set
with $m_c$ variations), taking fully into account all correlations
between $m_c$ and the PDFs.
In Fig.~\ref{fig:mtw-charm} we show the ratio of transverse mass
distributions computed with different $m_c$ in the PDFs,
divided by the results of the central NNPDF2.1 set.
It is clear from these results that a different choice of the charm mass
in the evolution of the parton densities
yields a different overall normalization of the transverse mass
distribution, but it affects very
moderately the  shape. This is confirmed by the
normalized distributions: the percentage difference with
respect to the reference $m_c$ value is consistent
with zero within statistical fluctuations.

In summary, we found in this section that PDF uncertainties in
the transverse mass distribution can be kept at the permille
level by normalizing them to the integrated cross section
in the fitted interval. These PDF uncertainties turn out to be
very similar for all colliders, energies and final states, and
are in reasonable agreement between different PDF sets.
The  theoretical uncertainties related to $m_c$ and 
$\alpha_s$, that are important for inclusive cross sections,
turn out to be negligible for the normalized kinematical distributions.
In the next section we will assess the impact of these various
uncertainties on the determination of $\mw$.

\section{PDF uncertainties in the determination of $\mw$ }
\label{resultsmw}

We have shown in the previous section how PDF uncertainties distort
the shape of the transverse mass distribution. Now we use the fit setup
presented in Sect.~\ref{fitting}  
to extract for each template the associated value of
$\mw$, and check how the values of $\mw$ obtained with
different PDF sets compare to each other and with their intrinsic
PDF uncertainties.
We fit the $W$ mass separately with each different Montecarlo replica
(for NNPDF2.1) or with the various Hessian eigenvectors (for
MSTW08 and CTEQ6.6), and then apply the corresponding prescriptions to
compute the best estimate for $\mw$ and the associated PDF uncertainty
for each set.
 
In Table~\ref{table:mwuncBorn-norm} we present the results obtained when 
fitting the Born level normalized transverse mass distributions 
(Eq.~\ref{eq:trans_mass_dist_norm}). 
Then in Table~\ref{table:mwuncBorn-unnorm} we show
the analogous results obtained when fitting the standard distributions 
(Eq.~\ref{eq:trans_mass_dist}).
 In both cases we quote the intrinsic PDF error from each set,
denoted by $\delta_{\rm pdf}$ (in GeV), as
well as the shift between each set and the reference value obtained
with CTEQ6.6, denoted by $\Delta_{\rm pdf}$ (in MeV).

We note that
the values of $\mw$ obtained with the standard
 distributions are shown here only for
illustration of the sensitivity of the template fit procedure to the
normalization choice.
The templates have been prepared at Born level, separately for each energy,
collider type and final state.
In this way we can claim that in each case we are probing only the effect
due to the PDF uncertainty.
We remark that the central value of CTEQ6.6 coincides, by
construction, with the  value ($\mw^0=80.398$ GeV) 
used when generating the pseudodata.

Let us consider first the results obtained with normalized
distributions, shown in Table~\ref{table:mwuncBorn-norm}.
The central values obtained with MSTW2008 and with NNPDF2.1
(that is the spread of $\Delta_{\rm pdf}$ values)
differ at most by 6 MeV with respect to $\mw^0$
and lie in general in a $\Delta_{\rm pdf} \sim 2-4$ MeV interval.
The PDF uncertainties are stable when considering different colliders,
energies and final states.
If we now look at the results obtained with the standard
distributions, Table~\ref{table:mwuncBorn-unnorm},
we observe that the central values are spread in a larger interval
($\pm 10$ MeV) and that also the PDF uncertainties are correspondingly
increased, $\delta_{\rm pdf} \sim 5-8$ MeV.  However, it is 
remarkable that even for the standard
transverse mass
distributions PDF uncertainties turn out to be rather small
and similar
for all colliders and final states.

\begin{table}[t]
\begin{center}
\small
\begin{tabular}{|c|c|c|c|c|c|c|}
\hline
collider,final state & \multicolumn{2}{c|}{CTEQ6.6}  &
\multicolumn{2}{c|}{MSTW2008}   & \multicolumn{2}{c|}{NNPDF2.1} \\
 & $\mw\pm \delta_{\rm pdf}$ & $\Delta_{\rm pdf}$ &
$\mw\pm \delta_{\rm pdf}$ & $\Delta_{\rm pdf}$  &$\mw\pm \delta_{\rm
  pdf}$ &$\Delta_{\rm pdf}$ \\
\hline
\hline
Tevatron, $W^\pm$ &  80.398  $\pm$ 0.004 & 0 &
                    80.399  $\pm$ 0.003 & +1 &
                   80.399  $\pm$ 0.005 & +1\\
\hline
LHC 7 TeV $W^+$ &  80.398  $\pm$ 0.003 & 0 &
                   80.404  $\pm$ 0.003 & +6 &
                   80.401  $\pm$ 0.003 & +3 \\
\hline
LHC 7 TeV $W^-$ &  80.398  $\pm$ 0.002 & 0 &
                   80.396  $\pm$ 0.002 & -2 &
                   80.400  $\pm$ 0.004 & +2 \\
\hline
LHC 14 TeV $W^+$ & 80.398  $\pm$ 0.003 & 0 &
                   80.402  $\pm$ 0.002  & +4 &
                   80.399  $\pm$ 0.003 & -1\\
\hline
LHC 14 TeV $W^-$ & 80.398  $\pm$ 0.002 & 0 &
                   80.398  $\pm$ 0.002 & 0 &
                    80.398  $\pm$ 0.005 & 0\\
\hline
\end{tabular}
\caption{\small \label{table:mwuncBorn-norm}
Results for 
the determination of $\mw$ from normalized transverse mass
 Born distributions.
We show in each case  the central value of the fit of $\mw$ and the spread
  due to  PDF uncertainties, $\delta_{\rm pdf}$ in GeV.
We also indicate 
well as $\Delta_{\rm pdf}$ (in MeV), the shift in central predictions
from each set  compared to the
CTEQ6.6 reference.}
\end{center}
\end{table}

\begin{table}[t]
\begin{center}
\small
\begin{tabular}{|c|c|c|c|c|c|c|}
\hline
collider,final state & \multicolumn{2}{c|}{CTEQ6.6}  &
\multicolumn{2}{c|}{MSTW2008}   & \multicolumn{2}{c|}{NNPDF2.1} \\
& $\mw\pm \delta_{\rm pdf}$ & $\Delta_{\rm pdf}$ &
$\mw\pm \delta_{\rm pdf}$ & $\Delta_{\rm pdf}$  &$\mw\pm \delta_{\rm
  pdf}$ &$\Delta_{\rm pdf}$ \\ 
\hline
\hline
Tevatron, $W^\pm$ & 80.398 $\pm$ 0.007 & 0 &
                   80.408 $\pm$ 0.007 & +10 &
                   80.407 $\pm$ 0.008 & +9 \\
\hline
LHC 7 TeV $W^+$ & 80.398 $\pm$ 0.007 & 0 &
                   80.399 $\pm$ 0.006 & +1 &
                   80.398 $\pm$ 0.005 & 0 \\
\hline
LHC 7 TeV $W^-$ & 80.398 $\pm$ 0.004 & 0 &
                   80.401 $\pm$ 0.004 & +3 &
                   80.399 $\pm$ 0.005 & +1 \\
\hline
LHC 14 TeV $W^+$ & 80.398 $\pm$ 0.008 & 0 &
                   80.393 $\pm$ 0.007 & -5 &
                   80.388 $\pm$ 0.005 & -10 \\
\hline
LHC 14 TeV $W^-$ & 80.398 $\pm$ 0.005 & 0 &
                   80.399 $\pm$ 0.004 & +1 &
                   80.391 $\pm$ 0.005 & -7 \\
\hline
\end{tabular}
\caption{\small
Same as Table~\ref{table:mwuncBorn-norm} for  $\mw$
fits to  the standard
transverse mass distributions.
\label{table:mwuncBorn-unnorm} }
\end{center}
\end{table}

\begin{figure}[t]
\begin{center}
\includegraphics[width=0.72\textwidth]{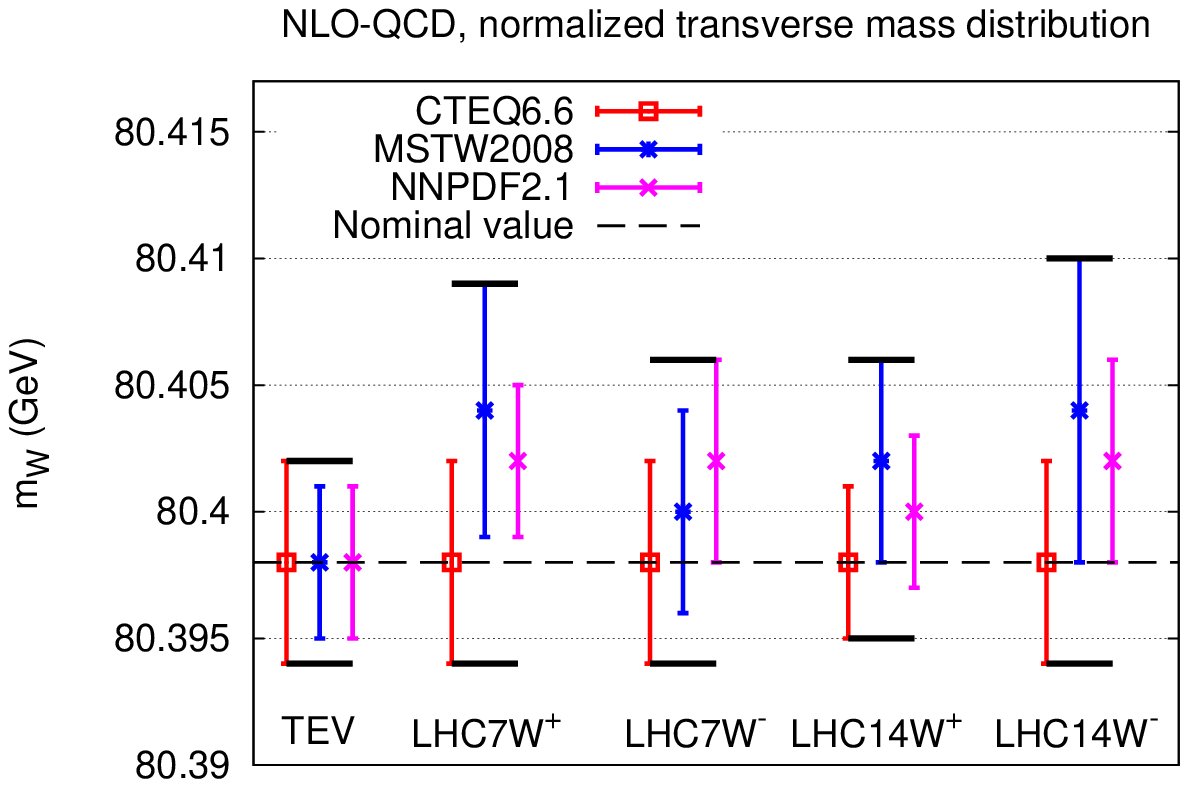}\\
\end{center}
\caption{\small Graphical representation of the results of 
Table~\ref{table:mwuncQCD} 
for the various colliders and final states considered. In each
case we draw the envelope of the results
from the PDF sets to define the total PDF uncertainty
($\pm\delta^{\rm tot}_{ \rm pdf}$) as a thick solid line. The dashed
line marks the position of the nominal value $\mw^0=80.398$ GeV used to
generate the pseudo-data.} 
\label{fig:mtw-plot}
\end{figure}

In Table~\ref{table:mwuncQCD} we present the results obtained by
fitting the transverse mass distribution generated at NLO-QCD with
DYNNLO, with different PDF sets.
The main difference with respect to the study at Born level
comes from the gluon contribution, absent in lowest order.
We fit the distributions, normalized to their cross-section in the
fitting interval, using DYNNLO templates prepared 
with the central CTEQ6.6 set and normalized. In 
Table~\ref{table:mwuncQCD} we also provide the average $\la\chi^2\ra$
obtained from the fit to the error PDFs
in each case. These results are represented graphically
in Fig.~\ref{fig:mtw-plot}.

The estimate of the PDF uncertainties for $\mw$ is quite stable
at different energies and colliders.
The values obtained here are moderately larger than at Born level.
To estimate the total PDF error, following
the PDF4LHC recommendation, we take the envelope of the
results from the three different PDFs sets. This
total PDF error obtained with the envelope method,
denoted by $\delta_{ \rm pdf}^{ \rm tot}$ in Table~\ref{table:mwuncQCD},
is always smaller than 10 MeV (see also Fig.~\ref{fig:mtw-plot}). Note in particular
the excellent agreement of the results from the three sets
at the Tevatron, both in central value and in PDF
uncertainty, yielding a total PDF uncertainty of only 
$\delta_{ \rm pdf}^{ \rm tot}=4$ MeV.

The above estimates of the PDF uncertainties are somewhat smaller
than previous estimates: for example, Ref.~\cite{atlasnote}
estimates $\delta_{ \rm pdf}^{\rm tot} \sim 25$ MeV prior to LHC
data.\footnote{ This uncertainty, computed from CTEQ6l, is to be
  understood as a 90\% C.L. and thus $\delta_{ \rm pdf}^{\rm tot} \sim
  15$ MeV at 68 \% C.L.}  We would like to emphasize that the key in
reducing the PDF uncertainties is fitting to the normalized kinematic
distributions, in a way that normalization effects in PDF
uncertainties, irrelevant for the $\mw$, cancel out. 
To illustrate this point, we note that from the results for $\mw$
obtained from the Born standard distributions,
Table~\ref{table:mwuncBorn-unnorm}, and using the envelope of the three
PDF sets, one finds $\delta_{ \rm pdf}^{ \rm tot} = 12$ MeV at Tevatron,  
$\delta_{ \rm pdf}^{ \rm tot} = 7(6)$ MeV at $W^+(W^-)$ LHC 7 TeV
and $\delta_{ \rm pdf}^{ \rm tot} = 12(9)$ MeV at $W^+(W^-)$ LHC 14 TeV,
larger than the results of Table~\ref{table:mwuncBorn-norm} 
(5 MeV, 6 MeV, 5 MeV, 5 MeV and 4 MeV respectively) and closer to previous estimates. 
Our estimates for PDF uncertainties on
$\mw$ at the Tevatron are also smaller than existing CDF and D0
estimates~\cite{:2009nu}. There is work in progress trying to understand
these various results and the differences
and similarities of the approaches.

\begin{table}[t]
\begin{center}
\small
\begin{tabular}{|c||c|c|l|c|l|c||c|}
\hline
  &  \multicolumn{2}{c|}{CTEQ6.6} &
\multicolumn{2}{c|}{MSTW2008}    & \multicolumn{2}{c|}{NNPDF2.1}  &\\
& $\mw\pm \delta_{\rm pdf}$& $\la \chi^2 \ra$&
$\mw\pm \delta_{\rm pdf}$& $\la \chi^2 \ra$ &
$\mw\pm \delta_{\rm pdf}$& $\la \chi^2 \ra$ &  $ \delta^{\rm tot}_{ \rm
  pdf}$ \\ 
\hline
\hline
Tevatron, $W^\pm$ & 80.398 $\pm$ 0.004 & 1.42 & 80.398 $\pm$ 0.003& 1.42
& 80.398 $\pm$ 0.003& 1.30  &  4 \\ 
\hline
LHC 7 TeV $W^+$ &   80.398 $\pm$ 0.004& 1.22 & 80.404 $\pm$ 0.005& 1.55
& 80.402 $\pm$ 0.003& 1.35 & 8 \\ 
\hline
LHC 7 TeV $W^-$ &   80.398 $\pm$ 0.004& 1.22 & 80.400 $\pm$ 0.004&1.19 &
80.402 
$\pm$ 0.004& 1.78 & 6 \\
\hline
LHC 14 TeV $W^+$ &  80.398 $\pm$ 0.003& 1.34 & 80.402 $\pm$ 0.004& 1.48
& 80.400 $\pm$ 0.003& 1.41 & 6\\ 
\hline
LHC 14 TeV $W^-$ &  80.398 $\pm$ 0.004& 1.44 & 80.404 $\pm$ 0.006& 1.38
& 80.402 $\pm$ 0.004& 1.57 & 8\\ 
\hline
\end{tabular}
\caption{\small Results for 
the determination of $\mw$ from normalized transverse mass
 NLO-QCD distributions.
We show in each case  the central value of the fit of $\mw$ and the spread
  due to  PDF uncertainties, $\delta_{\rm pdf}$ in GeV.
 In the right column of each PDF set, the average
  $\la\chi^2 \ra$ per degree of freedom obtained in the fit of the PDF 
error sets is shown, as a measure of the fit quality. For each collider
and final state, the final column estimates the total PDF uncertainty 
$\delta_{\rm pdf}^{\rm tot}$
using the envelope method, as discussed in the text. A graphical
representation of the results is shown in Fig.~\ref{fig:mtw-plot}.}
\label{table:mwuncQCD}
\end{center}
\end{table}

In Table~\ref{table:mwuncCTEQ45} we present the results obtained by
fitting the normalized transverse mass distributions, at the Tevatron,
with ResBos and with CTEQ6.6. The templates used in the fit have been
computed with ResBos as well, with the central CTEQ6.6 set.
Note that within the public version of ResBos only the CTEQ6.6 set
can be used.
By construction the central values of the fit coincide with the
nominal input value $\mw^0$.
The results for the PDF uncertainties are similar to those obtained with DYNNLO at
NLO-QCD.

\begin{table}[t]
\begin{center}
\small
\begin{tabular}{|c|c|c|}
\hline
collider,final state & CTEQ6.6\\
 & $\mw\pm \delta_{\rm pdf}$\\
\hline
Tevatron, $W^\pm$ & 80.398$\pm$0.006\\
\hline
\end{tabular}
\caption{\small 
Results for 
the determination of $\mw$ from normalized transverse mass
 NLO+NLL QCD distributions generated with ResBos.
We show the central value of the fit of $\mw$ and the spread
  due to  PDF uncertainties, $\delta_{\rm pdf}$ in GeV for
the case of the Tevatron.
  The distributions, normalized to the corresponding cross section in
  the fitting interval,
  have been computed using ResBos
  and have been fit with templates prepared as well with ResBos (with the
  central set of CTEQ6.6).}
\label{table:mwuncCTEQ45}
\end{center}
\end{table}

Let us consider now the impact of the values of $m_c$
and $\alpha_s$ on the determination of $\mw$, which we
know to be small from the analysis of the transverse
mass distributions.
In Table~\ref{table:mwuncALPHAS} we show the results found
in the determination of $\mw$ for the case of the NNPDF2.1
fits with varying strong coupling. As expected from
the distributions of Fig.~\ref{fig:mtw-alphas}, differences are
negligible, confirming that the uncertainty in $\alpha_s$
does not play a role for the determination of $\mw$ at
hadronic colliders.

\begin{table}[t]
\begin{center}
\small
\begin{tabular}{|c|c|c|c|c|c|c|c|c|c|c|}
\hline
        & Tevatron & LHC7W+ & LHC7W- & LHC14W+& LHC14W- \\
\hline
\hline
$\alpha_s(m_Z)=0.118$ & 80.398 & 80.400 & 80.398 & 80.402 & 80.400 \\
\hline
$\alpha_s(m_Z)=0.119$ (ref) & 80.398 & 80.402 & 80.402 & 80.400 & 80.402 \\
\hline
$\alpha_s(m_Z)=0.120$ & 80.398 & 80.400 & 80.398 & 80.402 & 80.402 \\
\hline
\end{tabular}
\caption{\small  Central value of the fit of $\mw$ 
obtained with NNPDF2.1, using PDF sets that differ by the $\alpha_s(m_Z)$ 
value,
for different colliders and energies.
The fit has been done on normalized distributions and using normalized
templates, and the distributions have been generated at NLO-QCD
with DYNNLO.
}
\label{table:mwuncALPHAS}
\end{center}
\end{table}

In Table~\ref{table:mwuncCHARM} we show the results obtained by
fitting the Born level
 transverse mass distributions generated with the NNPDF2.1
sets extracted with different values of $m_c$.
As it is evident from Fig.~\ref{fig:mtw-charm}, the results differ
mainly in their normalization.
Indeed we observe that the results of the Born level fit of the
normalized distributions
have very  small deviations with respect to the reference value
for the charm mass,
at most 4 MeV. This is to be compared with the sizeable shifts,
at the percent level, observed in the inclusive cross sections
when $m_c$ is varied in the global fit. Again the origin
of these different behaviors is that while cross sections depend
on the normalization of the distribution and thus of the
PDFs, $\mw$ depends only on its shape.

\begin{table}[t]
\begin{center}
\small
\begin{tabular}{|c|c|c|c|c|c|c|c|c|c|c|}
\hline
 $\mw$ (GeV)        & Tevatron & LHC7W+ & LHC7W- & LHC14W+& LHC14W- \\
\hline
\hline
$m_c=1.414$ (ref) & 80.398 & 80.402 & 80.402 & 80.400 & 80.402 \\
\hline
$m_c=1.5$ & 80.398 & 80.400 & 80.398 & 80.398 & 80.399 \\
\hline
$m_c=1.6$ & 80.398 & 80.400 & 80.400 & 80.398 & 80.399 \\
\hline
$m_c=1.7$ & 80.396 & 80.400 & 80.400 & 80.396 & 80.398 \\
\hline
\end{tabular}
\caption{\small  Central value of the fit of $\mw$ 
obtained with NNPDF2.1 sets with different values of
$m_c$
for different colliders and energies. We include the default
value in NNPDF2.1, $m_c^2=2 $ GeV$^2$ as well.
}
\label{table:mwuncCHARM}
\end{center}
\end{table}

\section{The impact of LHC data on the $\mw$ measurement}
\label{pdflhc}

In the previous sections we have discussed the impact of
PDF uncertainties on the determination of $\mw$ at hadron
colliders. The PDF sets considered there summarize our
understanding of the proton structure prior
to the LHC. However, LHC data is already providing stringent
constraints on available PDF, and thus it will further reduce
the effects of PDF errors in the extracted value for 
$\mw$. As an illustration of this
point, in this section we explore the impact that recent
ATLAS and CMS measurements of the lepton asymmetry from
$W$ decays have on the determination of $\mw$.

The CMS and ATLAS experiment have recently presented
their measurements of the lepton charge asymmetry from
$W$ bosons at 7 TeV. The ATLAS 
analysis~\cite{Aad:2011yn}  corresponds to
muon asymmetries while the CMS analysis~\cite{Collaboration:2011jz} 
contains both
the electron and the muon asymmetries for two cuts
of the lepton transverse momentum, $p_T\ge 25$ GeV
and $p_T\ge 30$ GeV. We have determined the impact of these
LHC measurements on the NNPDF2.1 set by means of
the Bayesian reweighting technique of Ref.~\cite{Ball:2010gb}.
Preliminary results were presented in~\cite{rojoPDF4LHC}, 
and a more detailed analysis
will be presented elsewhere.
For the comparison with LHC data, the theoretical predictions were
 generated at NLO with the 
DYNNLO event generator with the same binning and
cuts as in the respective experimental analysis.

We consider two cases: one in which the NNPDF2.1 set includes
the impact of the published CMS and ATLAS lepton asymmetry data
and another in which the NNPDF2.1 set includes the impact
of hypothetical future measurements of the same asymmetries with
a relative accuracy of 1\% (the average error of the published
data is about 7\%). 

\begin{figure}[t]
\begin{center}
\includegraphics[width=0.49\textwidth]{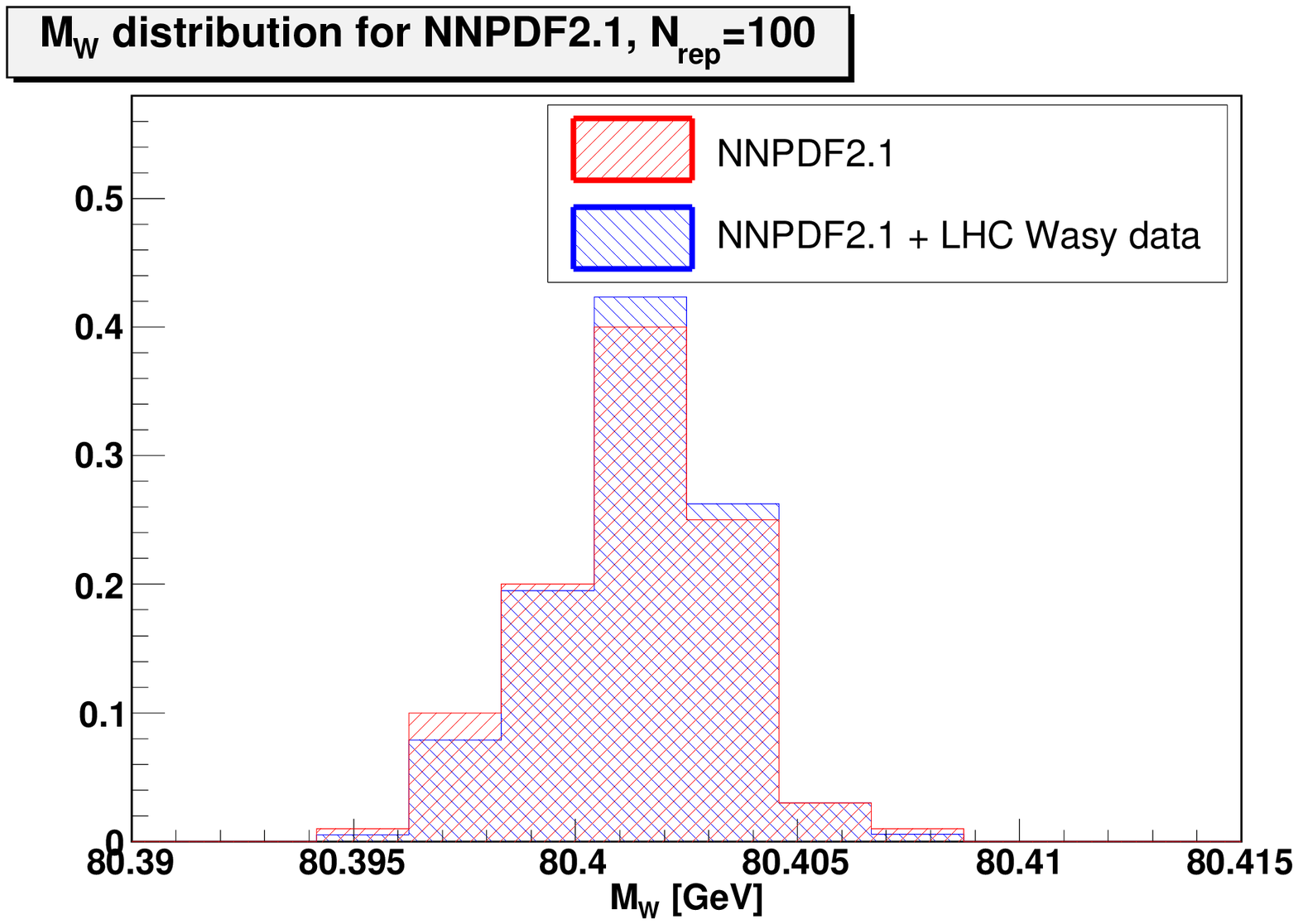}
\includegraphics[width=0.49\textwidth]{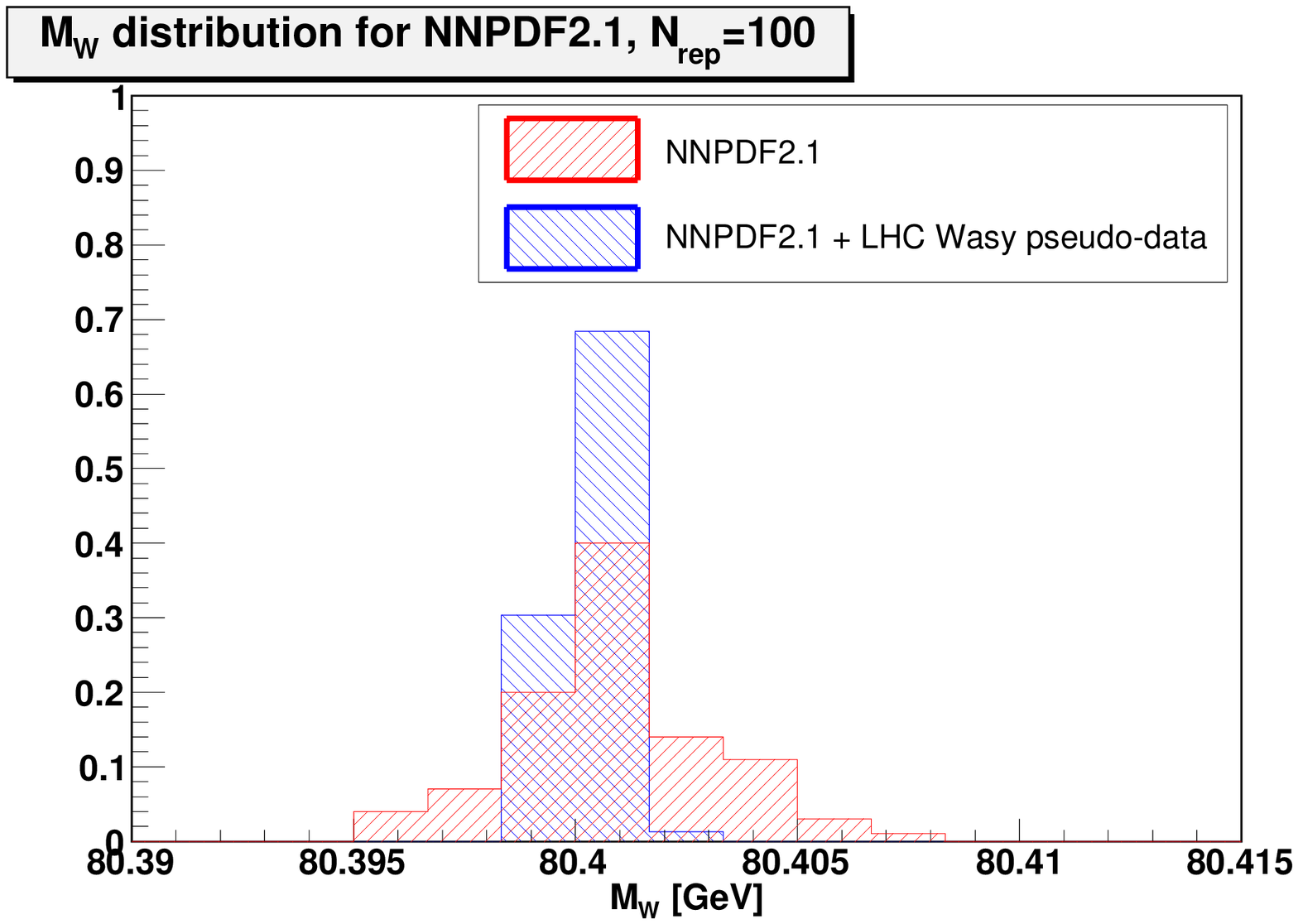}
\end{center}
\caption{\small The distribution of the 100 values of $\mw$ obtained
from each of the 100 replicas of the NNPDF2.1 analysis at the
LHC 7 TeV, compared
to the results of reweighting NNPDF2.1 with the recent ATLAS
and CMS data on $W$ lepton asymmetries (left plot) and by
the reweighting of $W$ lepton asymmetries pseudo-data
at the same kinematics than the published
LHC data but assuming with a 1\% total experimental uncertainty (right
plot). 
In each case the spread of the distributions represents the PDF
uncertainty.
See text for more details. } 
\label{fig:mwdist-nnpdf-lhc}
\end{figure}

We show the $\mw$ distributions obtained with the
$N_{\rm rep}=100$ replicas of NNPDF2.1 at the LHC 7 TeV
and the same  NNPDF2.1 replicas reweighted with the lepton asymmetry data 
in Fig.~\ref{fig:mwdist-nnpdf-lhc} (left plot).
The spread of the distribution indicates the PDF uncertainty
in the determination of  $\mw$.
 We can see that present
data already act in the direction of narrowing the distribution
thus reducing the PDF uncertainties in $\mw$, although the constraints
are still moderate. 

Larger effects are expected in the scenario
with lepton asymmetry pseudo-data with
a 1\% bin-per-bin total experimental uncertainty.  We can see  that
these very accurate pseudo-data have the potential to narrow
the $\mw$ distribution and thus to reduce the PDF
error on $\mw$ by a factor of two or even more. PDF uncertainties could
be further decreased if additional 
observables, sensitive to the quark and antiquark
combinations relevant for $\mw$ production, were considered.
One  example is provided by the accurate measurement of the
$Z$ boson rapidity distribution at the LHC, that would
constrain the small-$x$ sea quarks.

This exercise confirms that, though PDF uncertainties in the
determination of $\mw$ are already small, they can be further decreased
systematically by LHC measurements. 

\section{Conclusions}
\label{concl}

In this paper we have presented a detailed study of the impact of
PDF uncertainties on the accurate determination of the 
$W$ boson mass in hadronic collisions.
We have concentrated on the shape of the transverse mass distribution
and we have used a template fit technique to determine a preferred
$\mw$ value, isolating the PDF effects from other sources
of theoretical uncertainties.

Our main conclusions are the following:
\begin{itemize}
\item
The Born level study shows that the prediction of the central values and
of the PDF uncertainties agree between the different PDF sets
and are stable when comparing different colliders, energies and final
states. 
\item
The NLO-QCD study shows results analogous to the Born level case, with a
moderate increase of the PDF uncertainty induced by the
gluon initiated subprocesses.
\item
The use of accurate templates, prepared for each specific collider,
energy and final state, allows to disentangle the role of the PDFs,
while keeping fixed all the other input parameters.
\item
A sensible and more accurate
 fit of the $W$ mass can be obtained by studying the shape
of kinematical distributions, removing normalization effects which should not 
be explained in terms of $\mw$ shifts.
\item PDFs and related uncertainties ($\alpha_s$,$m_c$) are estimated
to be smaller than 10 MeV at the LHC for all energies and final
states, even before accounting from the improvements from LHC data.
This implies that PDF uncertainties will be smaller than other
systematic uncertainties.
\item PDF uncertainties, that are already rather moderate, can
be further reduced using LHC data alone, without the need of
a new dedicated experimental program to constrain PDFs.
We have illustrated this point using the recent lepton asymmetry
data from CMS and ATLAS. Measurements of the Z rapidity distribution
and other observables will soon further reduce PDF uncertainties.
Therefore a measurement at the level of 10 MeV precision
at the LHC, while challenging from many other points of view, does not
seem to be forbidden by the uncertainties in our knowledge of the proton
structure.
\end{itemize}

The precision determination of $\mw$ is one of the goals
of the current 7 TeV run at the LHC, due to its potential
to indirectly probe new physics at the electroweak scale.
This study ensures that an accuracy of 10 MeV is certainly
within reach, at least in what concerns our present knowledge
of the structure of the proton.

\vspace{0.5cm}
{\bf Acknowledgments}\\
We are grateful to M. Mangano for encouragement in
this project. We thank C. Carloni Calame, S. Forte, A. Kotwal, G. Montagna, O. Nicrosini and O. Stelzer-Chilton for stimulating discussions. We are grateful to M. Ubiali for providing
the DYNNLO predictions with NNPDF2.1 for the ATLAS and CMS lepton
asymmetry data discussed in Sect.~\ref{pdflhc}.

\end{document}